\newcommand{\blue}[1]{\textcolor{blue}{#1}}
\newcommand{\red}[1]{\textcolor{red}{#1}}

\newcommand\allbold[1]{{\boldmath\textbf{#1}}}

\documentclass[structabstract]{aa}  
\usepackage{graphicx}
\usepackage{lscape}
\usepackage{longtable}
\usepackage[]{natbib}
\usepackage{txfonts}
\usepackage{color}
\usepackage{url}
\usepackage{enumerate}
\newcommand{\mc}[3]{\multicolumn{#1}{#2}{#3}}

\begin{document}
%
\title{Optical polarisation variability of radio loud narrow line Seyfert 1 galaxies}
\subtitle{Search for long rotations of the polarisation plane}
\titlerunning{Optical polarisation variability of RL NLSy1 galaxies} 

\author{E.~Angelakis\inst{1}
  \and   
  S. Kiehlmann\inst{2}  
  \and   
  I. Myserlis\inst{1}  
  \and   
  D. Blinov\inst{3}  
  \and   
  J. Eggen\inst{4,5}  
  \and   
  R. Itoh\inst{6}  
  \and   
  N. Marchili\inst{7}  
  \and
  J.~A.~Zensus\inst{1}  
}

\institute{Max-Planck-Institut f\"ur Radioastronomie, Auf dem H\"ugel 69, 53121 Bonn, Germany\\
  	\email{eangelakis@mpifr-bonn.mpg.de}
  	\and
  	Owens Valley Radio Observatory, California Institute of Technology, Pasadena, CA 91125, USA
  	\and
	Department of Physics and Institute of Theoretical \& Computational Physics, University of Crete, GR-710 03, Heraklion, Greece  
	\and
  	Center for Research and Exploration in Space Science, NASA Goddard Space Flight Center, Greenbelt, MD 20771, USA
	\and
  	Department of Astronomy, University of Maryland, College Park, MD 20742, USA
  	\and
	Department of Physics, School of Science, Tokyo Institute of Technology, 2-12-1 Ohokayama, Meguro, Tokyo 152-8551, Japan  
  	\and
	Istituto di Astrofisica e Planetologia Spaziali	Via Fosso del Cavaliere 100, 00133, Rome, Italy
  }

\authorrunning{Angelakis et al.} 

\date{Received ; accepted}

 
  \abstract
   {Narrow line Seyfert 1 galaxies (NLSy1s) constitute the AGN subclass associated with systematically smaller black hole masses. A few radio loud ones have been detected in MeV -- GeV energy bands by \textit{Fermi} and evidence for the presence of blazar-like jets has been accumulated.  }
   {
      {In this study we wish to quantify the temporal behaviour of the optical polarisation, fraction and angle, for a selected sample of radio loud NLSy1s. We also search for rotations of the polarisation plane similar to the kinds of rotations that are commonly observed in blazars.} }
   {{We have conducted $R$-band optical linear polarisation monitoring of a sample of 10 radio loud narrow line Seyfert 1 galaxies; five of them being previously detected by \textit{Fermi}. The dataset obtained with our pivoting instrument: the RoboPol polarimeter of the Skinakas observatory, has been complemented with observations from the KANATA, Perkins and Steward observatories. In the cases where evidences for long rotations of the polarisation plane are found (at least three consecutive measurements covering at least $90\degr$), we carry out numerical simulations to assess the probability that they are caused by intrinsically evolving EVPAs instead of observational noise.} }
   {{Even our moderately sampled sources show clear indications of variability, both in polarisation fraction and angle. For the four best sampled objects in our sample we find multiple periods of significant polarisation angle variability. Several of these events qualify as long rotations. In the two best sampled cases, namely J1505$+$0326 and J0324$+$3410, we find indications for three long rotations of the polarisation angle. We show that although noise can induce the observed behaviour, it is much more likely that the apparent rotation is indeed caused by intrinsic evolution of the EVPA. To our knowledge this is the very first detection of such events in this class of sources. In the case of the largest dataset (J0324$+$3410) we find that the EVPA concentrates around a direction which is at $49.3\degr$ to the 15-GHz radio jet implying a projected magnetic field at an angle of $40.7\degr$ to that axis. 
   }}
   {{We assess the probability that pure measurement uncertainties are the reason behind the observed long rotations of the polarisation plane. We conclude that although this is not improbable, it is much more likely that intrinsic rotations are responsible for the observed phenomenology. We conclude however that much better sampled and larger datasets of larger source samples are necessary constraining the physical mechanism(s) that generate long EVPA rotations in NLSy1s.}}

   \keywords{Galaxies: active -- Galaxies: Seyfert -- Polarization -- Methods: numerical -- Methods: statistical -- Techniques: polarimetric}

   \maketitle

\section{Introduction}
\label{sec:introduction}

The term narrow-line Seyfert 1 galaxies (hereafter NLSy1s) signifies the subset of AGN with narrow width of the broad Balmer emission line (FWHM(H$\beta$) $\lesssim 2000$~km~s$^{-1}$), and weak forbidden lines with [\ion{O}{iii}]$\lambda$5007/H$\beta$ $<3$ \citep[][]{1985ApJ..297...166,1989ApJ...342..224G,2006ApJS..166..128Z}. They are thus associated with black hole masses in the range $10^{6}$--$10^{8}$~M$_{\odot}$ \citep[e.g.][]{2012AJ....143...83X,2015A&A...575A..13F}  smaller than those of powerful radio galaxies that typically exceed $10^{8}$~M$_{\odot}$. Assuming these estimates to be free of biases \citep[for claims of the opposite see][]{2008ApJ...678..693M,2016MNRAS.458L..69B}, the detection of jet GeV emission \citep{2009ApJ...707L.142A,2012MNRAS.426..317D,2009ApJ...699..976A,2015MNRAS.452..520D} and jet radio emission \citep[e.g.][]{2012A&A...548A.106F,2015A&A...575A..55A,2017A&A...603A.100L} from radio-loud (RL)\footnote{the radio loudness $R$ is defined as the ratio of the 6~cm flux to the optical flux at 4400~$\AA$  \citep{Kellermann1989AJ}} NLSy1s, challenges the current understanding of relativistic jet formation in which {powerful relativistic jets are preferentially found in elliptical galaxies} with nuclear black hole masses beyond $10^{8}$~M$_{\odot}$ \citep[for a review of the main arguments see e.g.][]{2017MNRAS.469L..11D}. 

In \cite{2015A&A...575A..55A} we presented a comprehensive study of the radio emission of the four RL NLSy1s detected by \textit{Fermi}. The dataset covered the band from 2.64~GHz to 142.33~GHz at 10 frequencies and with a cadence of less than 30 days (for three of the four sources). Despite the generally lower fluxes, all sources showed typical {characteristics seen in blazars: intense variability accompanied by dramatic spectral evolution indicative of shocks operating in a plasma outflow}. We computed limiting values of the brightness temperature and inferred rather moderate Doppler factors implying the presence of mildly relativistic jets. The computed jet powers appeared comparable to the least energetic blazars, the BL Lac objects. In conclusion, the sources showed all the typical characteristics of blazars only scaled to lower intensities. In \cite{2016RAA....16..176F} we focused on the dynamics of the 15~GHz jet of 1H\,0323$+$342. Our analysis revealed superluminal components indicative of a relativistic jet, from which we inferred a viewing angle of less than $9^\circ$ confirming the ``aligned jet'' scenario.    

We currently focus on the optical polarisation of RL~NLSy1s. \cite{2012AAS...21924329E} and \cite{2013ApJ...773...85E}, among the first studies on the subject, reported that PMN\,J0948$+$0022 showed significant and variable polarisation in optical bands. The same source was observed by \cite{2013ApJ...775L..26I} who found minute time scale optical polarisation variability. During this ``pulse'' the polarisation exceeded  30~\% while, interestingly, the polarisation angle i.e. the electric vector position angle (EVPA), appeared unchanged. The authors interpreted their findings as evidence of synchrotron emission radiated from a compact region of  highly ordered magnetic field. In the case of J0849$+$5108 on the other hand, \cite{2014ApJ...794...93M} observed rapid intra-night variability in polarisation degree and angle during a major broadband outburst event which lasted for roughly five days. More recently, \cite{2014PASJ...66..108I} studied 1H\,0323$+$342. They reported that the EVPA remained {roughly} parallel to the jet orientation implying a magnetic field transverse to the jet axis. 

Long rotations of the optical polarisation plane have been found in blazars \citep[e.g.][]{1988A&A...190L...8K,2008Natur.452..966M,2010ApJ...710L.126M,2010Natur.463..919A,2016MNRAS.462.1775B}. Models that have been put forth to interpret the observations include: physical rotation of emission elements on a helical trajectory \citep{2008Natur.452..966M}, propagation in large {a} scale bent jet \citep{2010Natur.463..919A}, turbulent plasma processes resulting {in} random walks \citep{2014ApJ...780...87M}, or light travel time effects within an axisymmetric emission region \citep{2015ApJ...804...58Z}. Interestingly, it has been argued {that these physical processes are likely} associated with {increased episodic} gamma-ray activity \citep{2018MNRAS.474.1296B}. 

Beyond the potential {of using} polarisation monitoring to probe the physical processes at the emission site, {the variability of the EVPA in particular} {can further} our understanding of the conditions {present} during the high energy jet emission production. In this context we wish to: {(a) quantify the variability of the $R$-band optical polarisation fraction and angle for a selected sample of RL NLSy1 galaxies, (b) examine whether long rotations of the polarisation plane occur in RL~NLSy1s, (c) parametrise {these rotations} and examine their association with the high energy activity, and (d) ultimately understand the {physical mechanisms} producing them.} 

Here we present a study of a sample of 10 RL~NLSy1s; five of them detected by \textit{Fermi} (c.f. Section~\ref{sec:observations} and table~\ref{tab:sample}). {Whenever the datasets allow us, we study the variability of both polarisation parameters and search for EVPA rotation candidates}. For the two best sampled cases {we asses the probability of these events being driven by intrinsic EVPA variability rather than observational noise. This distinction is accomplished by conducting exhaustive simulations. In the following we emphasise {both our} findings as well as the method for assessing the probability itself. In these two cases we indeed find evidence that the rotations are intrinsic to each source. } This is the first time that such events are reported for RL~NLSy1s.


\section{Source sample and dataset}
\label{sec:observations}

The selection of our sample has been based mostly on the radio loudness (RL) and the observability of the sources from the Skinakas telescope (i.e. optical magnitude and position).  
It includes five of the eight sources that have been reported to radiate significant emission in the MeV -- GeV energy range \citep{2009ApJ...699..976A,2009ApJ...707L.142A,2012MNRAS.426..317D,2015MNRAS.452..520D,2015MNRAS.454L..16Y,2015arXiv151005584L}. 
It also includes another five RL sources that make up a total of ten targets. All the sources with at least one data point from our monitoring have been listed in Table~\ref{tab:sample}. {Median values and ranges of the polarisation parameters for all the sources discussed here are shown in Table~\ref{tbl:cumm_pol}}.    
\begin{table*}[]
  \caption{List of sources in our sample and their relevant parameters.}
  \label{tab:sample}  
  \centering                    
  \begin{tabular}{llcr@{\hskip 0.02cm}lrl} 
    \hline\hline                 
ID            &Survey ID                    &\mc{1}{l}{Redshift}            &\mc{2}{c}{$M_\mathrm{BH}$}  &\mc{1}{c}{$R$}    &Notes  \\
    \hline
\\                                                                                                                       
J0324$+$3410  & 1H\,0323$+$342              &0.062900 $^\mathrm{1}$ &$2   - 3.4$ &$\times10^{7~\mathrm{A, P, Q}}$  & 318 $^\mathrm{O}$    &\parbox[][][b]{5cm}{Fermi detected$^\mathrm{5}$} \\
J0849$+$5108  & SBS\,0846$+$513             &0.584701 $^\mathrm{2}$ &$0.8 - 9.8$ &$\times10^{7~\mathrm{B, C, D  }}$   &1445 $^\mathrm{J}$    &\parbox[][][b]{5cm}{Fermi detected$^\mathrm{6}$} \\ 
J0948$+$0022  & PMN\,J0948$+$0022           &0.585102 $^\mathrm{2}$ &$0.2 - 8.1$ &$\times10^{8~\mathrm{E, F,10  }}$   & 355 $^\mathrm{J,10}$ &\parbox[][][b]{5cm}{Fermi detected$^\mathrm{7}$} \\
J1305$+$5116  & WISE J130522.75$+$511640.3  &0.787552 $^\mathrm{2}$ &$3.2      $ &$\times10^{8~\mathrm{J        }}$   & 223 $^\mathrm{J}$    &\parbox[][][b]{5cm}{Optical spec. indicates strong outflow$^\mathrm{11}$.}            \\
J1505$+$0326  & PKS\,1502$+$036             &0.407882 $^\mathrm{2}$ &$0.04 - 2 $ &$\times10^{8~\mathrm{G,H, 5, I,11}}$&1549 $^\mathrm{J}$    &\parbox[][][b]{5cm}{Fermi detected$^\mathrm{5}$} \\
J1548$+$3511  & HB89\,1546$+$353            &0.479014 $^\mathrm{2}$ &$7.9      $ &$\times10^{7~\mathrm{J        }}$   & 692 $^\mathrm{J}$    &\parbox[][][b]{5cm}{Evidence for past radio variability.}            \\
J1628$+$4007  & RX\,J16290$+$4007           &0.272486 $^\mathrm{2}$ &$3.5      $ &$\times10^{7~\mathrm{L,10     }}$   &  29 $^\mathrm{N,10}$ &\parbox[][][b]{5cm}{High optical and radio variability$^\mathrm{10}$.}      \\
J1633$+$4718  & RX\,J1633.3$+$4718          &0.116030 $^\mathrm{4}$ &$3        $ &$\times10^{6~\mathrm{K        }}$   & 166 $^\mathrm{J}$    &\parbox[][][b]{5cm}{Evidence for past radio variability.}            \\
J1644$+$2619  & FBQS\,J1644$+$2619          &0.145000 $^\mathrm{3}$ &$2.1      $ &$\times10^{8~\mathrm{M        }}$   & 447 $^\mathrm{N}$    &\parbox[][][b]{5cm}{Fermi detected$^\mathrm{8}$}         \\
J1722$+$5654  & SDSS\,J172206.02$+$565451.6 &0.425967 $^\mathrm{2}$ &$2.5 - 3.3$ &$\times10^{7~\mathrm{J,9      }}$   & 234 $^\mathrm{J,9}$  &\parbox[][][b]{6cm}{Evidence for high-amplitude  optical variability$^\mathrm{9}$.} \\
\\    \hline                                  
  \end{tabular}
  \tablefoot{Columns: (1) Source identifier, (2) Survey identifier, (3) redshift, (4) black hole mass, (5) radio loudness. 
   \tablebib{(1) \cite{2007ApJ...658L..13Z};  (2) \cite{2010MNRAS.405.2302H}; (3) \cite{2015A&A...575A..13F}; (4) \cite{2015ApJS..219....1O}; (5) \cite{2009ApJ...707L.142A}; (6) \cite{2012MNRAS.426..317D}; (7) \cite{2009ApJ...699..976A}; (8) \cite{2015MNRAS.452..520D}; (9) \cite{2006ApJ...639..710K}; (10) \cite{2006AJ....132..531K}; (11) \cite{2016IAUS..312...63K}; (A) \cite{2017MNRAS.464.2565L}; (B) \cite{2005ChJAA...5...41Z}; (C) \cite{2011ApJS..194...45S}; (D) \cite{2016ApJ...819..121P}; (E) \cite{2003ApJ...584..147Z}; (F) \cite{2009ApJ...699..976A}; (G) \cite{2008ApJ...685..801Y}; (H) \cite{2016ApJ...820...52P}; (I) \cite{2013MNRAS.431..210C}; (J) \cite{2008ApJ...685..801Y}; (K) \cite{2010ApJ...723..508Y}; (L) \cite{2015A&A...575A..13F}; (M) \cite{2017MNRAS.469L..11D}; (N) \cite{2016PASJ...68...73D}; (O) \cite{2011nlsg.confE..24F} ; (P) \cite{2016ApJ...824..149W} ; (Q) \cite{2018MNRAS.475..404K} }
   }
\end{table*}



\subsection{The RoboPol dataset}
\label{sub:robopol}
The RoboPol\footnote{http://robopol.org} dataset has been the basis for the present study. The instrument is mounted on the 1.3-m telescope of the Skinakas observatory \citep{2007Ippa....2b..14P} and has been monitoring our sample in the $R$-band. All the details of the measurement techniques and the instrument characteristics are discussed in \cite{2014MNRAS.442.1706K} and \cite{2016MNRAS.463.3365A} where post-measurement quality criteria are discussed in detail. 

\subsection{The KANATA dataset}
\label{sub:kanata}
The Kanata observations were conducted with the 1.5-m telescope of Higashi-Hiroshima Observatory. 
The polarimetry was performed with the HOWPol polarimeter \citep{2008SPIE.7014E..4LK}. The observing cycle includes successive exposures at four position angles of a half-wave plate at 0, 45, 22.5, and $67.5^\circ$. The instrumental polarisation (peaking at $\sim4~\%$), was modelled and removed before further analysis. The residual uncertainties are estimated from large numbers of unpolarised standard stars, and is smaller than 0.5~\%.

\subsection{The Perkins dataset}
\label{sub:perkins}
The Perkins dataset was obtained with a Johnson $R$ filter using the PRISM instrument on the 1.8-m Perkins Telescope of the Lowell Observatory which also includes a rotating half-wave plate polarimeter. 
The observing cycle included exposures with the half-wave plate at 0, 45, 90, and $135^\circ$. The averages of two to four such cycles was used as the final measurement. 
Instrumental offsets of the EVPA and percent polarisation (usually less than 1\%) were determined by observing in-field polarised and unpolarised standard stars \citep{1992AJ....104.1563S}.


\subsection{The Steward Observatory dataset}
\label{sub:steward}

The Steward Observatory data have also been obtained at $R$-band and they have been retrieved from the online archive\footnote{http://james.as.arizona.edu/\~{}psmith/Fermi/}. The data acquisition and reduction is described in \cite{2009arXiv0912.3621S}.


\section{Rice bias treatment}
\label{sec:bias}

The functional dependence of the polarisation fraction $p$ on the normalised Stokes parameters $q$ and $u$ introduces a bias in its determination from repeated observations in the presence of noise. The effect becomes particularly important at low signal-to-noise ratios (SNRs). Here we study the temporal behaviour of the EVPA from observations of already moderate sampling. To avoid unnecessary data loss that could be imposed by using only high significance data we use all available observations after we treat the Rice bias. 

Let us assume a fixed polarisation vector with real amplitude $p_0$ at an angle of $\chi_0$ which is observed in the presence of experimental noise. Unless the SNR is  large, none of the observed polarisation parameters $p$ and $\chi$ determined from repeated observations will follow a normal distribution even though $\chi$ will {populate} a distribution symmetric around $\chi_0$. 

Due to the presence of Gaussian noise the $q$ and $u$ will be normally distributed about their true values $q_0$ and $u_0$ respectively and with equal uncertainties $\sigma_q=\sigma_u=\sigma_{q_0}=\sigma_{u_0}$ which are also equal to {the uncertainties} in $p$, $\sigma_p$. The probability, however, of measuring polarisation in the range $\left[p,p+dp\right]$ {independent} of polarisation angle ({integrating over all angles}) -- as it was first demonstrated by \cite{1958AcA.....8..135S} -- will be given by the Rice distribution \citep{1945BSTJ...24...46R}
\begin{equation}
	\label{eq:rice}
	F(p \mid p_0)dp=\frac{p}{\sigma_p}\exp{\left[-\frac{p^2+p_0^2}{2\sigma_p^2}\right]}I_0\left(\frac{pp_0}{\sigma_p^2}\right)\frac{dp}{\sigma_p}
\end{equation}
with $I_0$ the zeroth-order modified Bessel function. The asymmetry of Eq.~\ref{eq:rice} with respect to $p$ and $p_0$ is the cause for the observed polarisation bias especially at low SNRs. At high SNRs the Rice distribution tends to a normal one with a mean around the true value of polarisation $p_0$ and a spread $\sigma_p$. \cite{1965AnAp...28..412V,1985A&A...142..100S,1993A&A...274..968N} have investigated the distributions of the observed amplitudes and the angles from repeated observation and thorough descriptions of the problem can be found in \cite{1974ApJ...194..249W,2006PASP..118.1340V, 2010stpo.book.....C}.    

Concerning the amplitude of polarisation we adopt the same approach as in \cite{2014MNRAS.442.1693P}; as a best-guess of $p_0$ we take the approximation of the maximum-likelihood estimator $\hat{p}$ given by \cite{2006PASP..118.1340V}: 
\begin{equation}
\hat{p} = \left\{ \begin{array}{lll}
 0 & &\mbox{for $p/\sigma_p<\sqrt{2}$} \\
  \sqrt{p^2-\sigma_p^2} & &\mbox{for $p/\sigma_p\ge \sqrt{2}$}
       \end{array} \right.
\end{equation}
The uncertainty in the de-biased polarisation fraction is set to the observed one $\sigma_{p}$ as long as it is bounded at zero.

Because we are interested in the occurrence of long rotations of the polarisation plane, it is particularly important to assess the uncertainty in the polarisation angle even in cases of low SNR. 
For that matter we adopt the approach presented by \cite{1993A&A...274..968N}. We solve their Eq.~4 for $\sigma_\mathrm{\theta}$ with  the SNR of the de-biased value ($\hat{p}/\sigma_p$). $\sigma_\theta$ is then taken as the uncertainty in the angle $\sigma_\chi$.

\section{Long rotations of the polarisation plane}
\label{sec:evpa_monecla}

In Sect.~\ref{sec:bias} we discussed the treatment of the polarisation fraction. Here we {clarify} the conventions and terminology used for the EVPA. 

The EVPAs are initially computed from the observed $q$ and $u$ as:
\begin{equation}
	\label{eq:chi}
\chi_\mathrm{obs}=\frac{1}{2}\cdot \arctan\left(\frac{u}{q}\right)
\end{equation}
and hence carry the {inherent} ``$n\times\pi$" ambiguity. For each $(q,u)$ pair we choose Eq.~\ref{eq:chi}'s solution for which the difference from the previous data point is less than $90^\circ$.  
The resulting data points are then termed \textit{adjusted} {and are the values we use in the following discussion}. This adjustment is made under the assumption of minimal variability between adjacent data points.

Phases of significant EVPA \textit{variability} are defined as sequences of data points over {which the adjusted polarisation angle $\chi$ changes significantly between consecutive observations} \allbold{($|\Delta\chi|>\sqrt{\sigma_{\chi_\mathrm{i}}^2+\sigma_{\chi_\mathrm{i+1}}^2}$)}. 
 Such periods are marked in plots as Fig.~\ref{fig:1505_t_p_evpa_long} by \textit{dotted} coloured lines. Lines of the same colour connect data points that show an \textit{overall} trend in the same direction. Over such periods \textit{insignificant} changes in the opposite direction are allowed. This approach is described in detail in \cite{2016A&A...592C...1K}. We {define} such periods as \textit{long} EVPA rotations {(i.e. long rotations of the polarisation plane)}, when they: 
\begin{enumerate}
\item consist of at least 3 data points, and  
\item exceed $90^\circ$.
\end{enumerate}
Such periods are marked with \textit{solid} coloured lines. A rotation is terminated when significant variability shows a change in sign.

\section{Analysis of the polarisation fraction and angle variability}
\label{sec:temporal} 



{Here}, we study the temporal behaviour of the optical polarisation (fraction and angle) for sources with large enough datasets. For the remaining we list all their RoboPol measurements in Table~\ref{tbl:debiased_data}. {Median values and ranges of the polarisation parameters for all the sources discussed here are shown in Table~\ref{tbl:cumm_pol}}. 
\begin{table}[h]
\caption{The RoboPol measurements for the sources that did not allow studies of the temporal evolution of the EVPA. The data have been corrected for the Rice bias.}
\label{tbl:debiased_data} 
\centering
\begin{tabular}{lrrrr}
\hline\hline
JD        &\mc{1}{c}{$\hat{p}$}    &\mc{1}{c}{$\sigma_{p}$}   &\mc{1}{c}{$\chi$}       &\mc{1}{c}{$\sigma_{\chi}$} \\
          &      &                 &\mc{1}{c}{($\degr$)}     &\mc{1}{c}{($\degr$)}       \\
\hline\\
\mc{5}{c}{J1305$+$5116} \\\\
 2457209.285  &  0.011  &  0.008  &   -29.6  &    21.2 \\
 2457240.322  &  0.008  &  0.007  &    -8.9  &    23.7 \\\\
\mc{5}{c}{J1548$+$3511} \\\\
 2457209.367  &  0.000  &  0.013  &    47.9  &    61.4 \\
 2457240.364  &  0.021  &  0.011  &   -10.8  &    16.5 \\
 2457264.291  &  0.058  &  0.016  &   -32.9  &     8.0 \\\\
\mc{5}{c}{J1628$+$4007} \\\\
 2457209.405  &  0.000  &  0.008  &   -25.0  &    61.4 \\
 2457254.323  &  0.000  &  0.009  &   -45.3  &    61.4 \\\\
\mc{5}{c}{J1633$+$4718} \\\\
 2457209.425  &  0.021  &  0.005  &    -8.1  &     7.0 \\
 2457228.395  &  0.030  &  0.005  &    -5.6  &     5.2 \\
 2457240.384  &  0.019  &  0.006  &     3.7  &     9.2 \\
 2457254.345  &  0.027  &  0.005  &    -3.1  &     5.0 \\\\
\mc{5}{c}{J1644$+$2619} \\\\
 2457209.445  &  0.039  &  0.008  &   -24.2  &     5.8 \\
 2457230.387  &  0.000  &  0.011  &   -32.5  &    61.4 \\
 2457240.404  &  0.031  &  0.006  &   -16.8  &     6.0 \\
 2457254.358  &  0.012  &  0.012  &   -34.8  &    28.8 \\\\
\mc{5}{c}{J1722$+$5654} \\\\
 2457228.434  &  0.000  &  0.014  &   -33.2  &    61.4 \\
 2457240.429  &  0.000  &  0.015  &   -37.7  &    61.4 \\\\
\hline
\end{tabular}
  \tablefoot{Columns: (1) Julian date, (2) de-biased polarisation fraction, (3) uncertainty in the de-biased polarisation fraction, (4) polarisation angle (EVPA), (5) uncertainty in the EVPA. 
   }
\end{table}
\begin{table}[h]
\caption{{Integrated polarisation characteristics.}}
\label{tbl:cumm_pol} 
\centering
\begin{tabular}{lrrrrrr}
\hline\hline
Source    &\mc{1}{c}{N} &\mc{1}{c}{$\left<\hat{p}\right>$} &\mc{1}{c}{$\sigma_{p}$}   &\mc{1}{c}{$\left<\chi\right>$} &\mc{1}{c}{$\chi_\mathrm{min}$} &\mc{1}{c}{$\chi_\mathrm{max}$}    \\
          &             &                                  &\mc{1}{c}{($\degr$)}      &\mc{1}{c}{($\degr$)}           &\mc{1}{c}{($\degr$)}           &\mc{1}{c}{($\degr$)}              \\
\hline\\
J0324        &115 &0.012   &0.016   &$-6.7$  &$-89.1$  &$+87.0$\\
J0849        &15  &0.100   &0.078   &33.3    &$+6.8  $ &$+62.6$\\
J0948        &30  &0.024   &0.028   &9.0     &$-83.0$  &$+79.7$\\
J1505        &26  &0.040   &0.030   &$-1.9$  &$-61.7$  &$+88.6$\\
J1305$+$5116 &2   &0.010   &0.002   &$-19.3$ &$-29.6$  &$-8.9$ \\
J1548$+$3511 &3   &0.021   &0.024   &$-10.8$ &$-32.9$  &$+47.9$ \\
J1628$+$4007 &2   &0.000   &\ldots  &\ldots  &\ldots   &\ldots   \\
J1633$+$4718 &4   &0.024   &0.004   &$-4.4$  &$-8.1 $  &$+3.7$  \\
J1644$+$2619 &4   &0.022   &0.015   &$-28.4$ &$-34.8$  &$-16.8$\\
J1722$+$5654 &2   &0.000   &\ldots  &\ldots  &\ldots   &\ldots  \\
\hline
\end{tabular}
{
  \tablefoot{Columns: (1) Source ID, (2) number of available measurements, (3) median de-biased polarisation fraction, (4) spread of polarisation fraction, (5) median polarisation angle,(6,7) min and max polarisation angle in the range $\left[-90\degr, 90\degr\right]$. 
   }
   }
\end{table}

\subsection{J1505$+$0326}
\label{subsec:J1505}

{We start with} the source J1505$+$0326, because (a) it shows clearly discernible events making {the implementation of our approach}  easier, and (b) because {our analysis shows} that it is the best candidate to have undergone an intrinsic long rotation {of the polarisation plane}. 

\begin{figure*}[h] 
\centering
\includegraphics[trim={0 0 0 0},clip, width=0.9\textwidth]{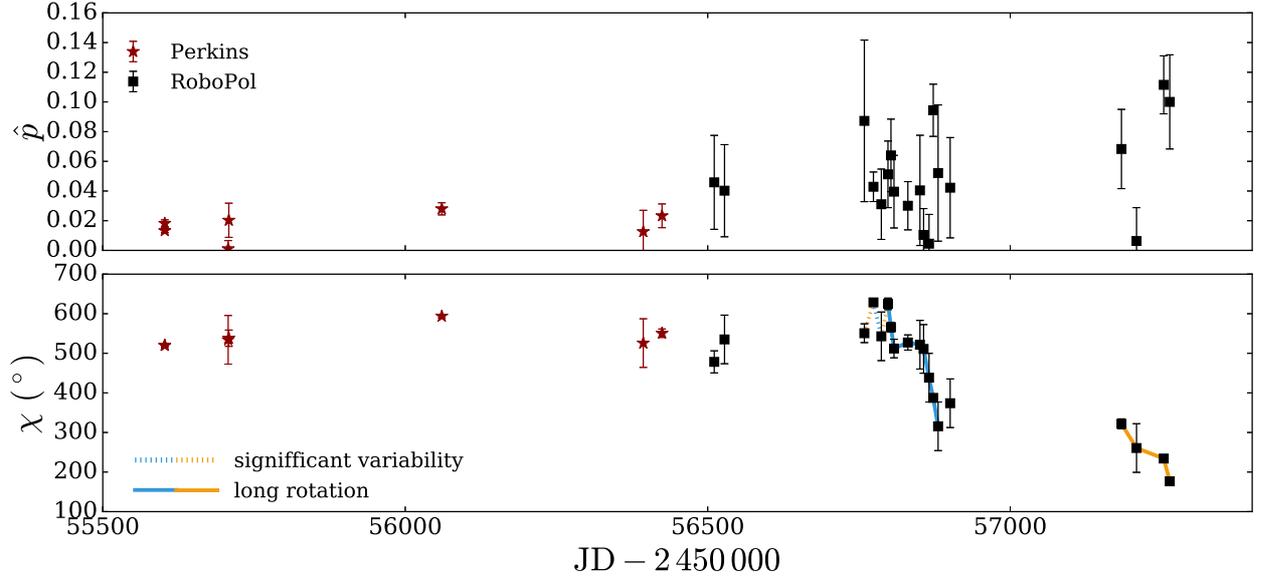} 
\caption{J1505$+$0326: De-biased polarisation fraction $p$ and adjusted EVPA ($\chi$) as a function of time. The coloured lines mark periods of significant monotonous -- within the uncertainties -- EVPA evolution. 
Solid lines mark periods of long rotations (i.e. at least three sequential data points and angle larger than $90^\circ$). {\textit{Blue}} and {\textit{orange}} connecting lines are used alternatively to ease reading.}
\label{fig:1505_t_p_evpa_long}
\end{figure*}

Figure~\ref{fig:1505_t_p_evpa_long} presents the RoboPol and Perkins observations of the optical linear polarisation parameters: $\hat{p}$ and $\chi$. 
The coloured lines there (dotted or solid) mark five periods of significant, continuous EVPA variability that we have detected. {The absolute rotation angles of those events -- in order of occurrence -- are 77.9,   85.7,   82.2,  309.5 and  $145.1^\circ$.}
{The three first events consist} of only two consecutive data points (dotted lines). The {last two however are made of at least three} sequential data points and exceed the limit of $90^\circ$; {and thus} qualify as {long rotation candidates} (solid lines). 

\begin{figure}[h] 
\centering
\includegraphics[trim={0 0 0 0},clip, width=0.5\textwidth]{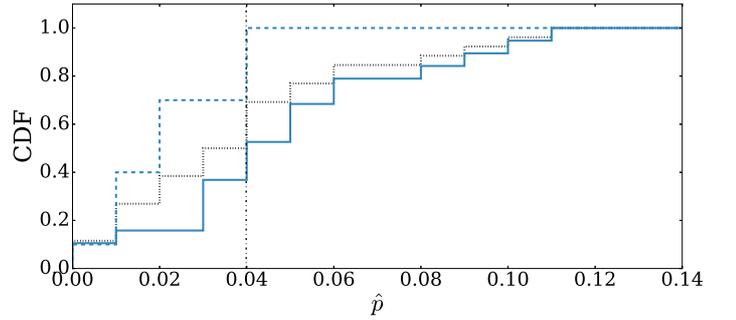} 
\caption{{J1505$+$0326: The distribution of the de-biased polarisation fraction $p$. The dot-dashed vertical line marks the median of the distribution. The blue dashed line shows the distribution of $\hat{p}$ during the non-rotating phases and the solid one that during the rotating phases. The black dotted line corresponds to all the measurements. } }
\label{fig:J1505_pol_hist}
\end{figure}

{The polarisation fraction spreads {around} a median of around 0.04 with a standard deviation $\sim0.03$ (Table~\ref{tbl:cumm_pol}). Figure~\ref{fig:J1505_pol_hist} shows its cumulative distribution function  for all the measurements (dotted black line), during phases of rotation (solid blue line) and during non-rotating phases (dashed blue line). The median, $\hat{p}$, of the non-rotating phases alone is 0.022 while that over rotating ones 0.043. Despite the difference, a two-sample KS test, however, gave no indication for different parent populations which in turn prohibits any conclusion about the behaviour of the polarisation during the rotations.}

{In the following we focus on the largest long rotation and asses the probability that it is driven by an intrinsically rotating polarisation plane.}


\subsubsection{J1505$+$0326: The long rotation}
\label{subsubsec:rotation}
\begin{figure*}[h] 
\centering
\includegraphics[trim={0 0 0 0},clip, width=0.9\textwidth]{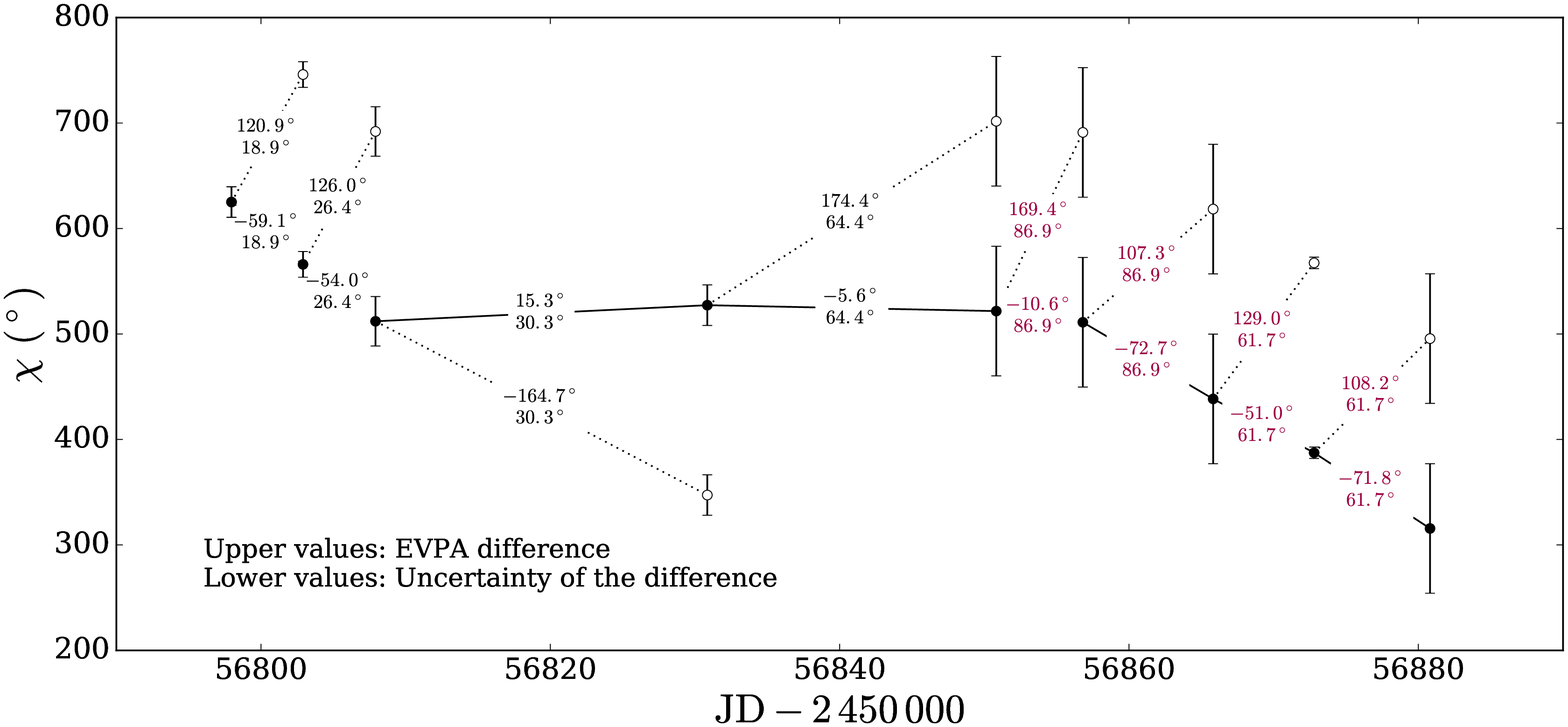} 
\caption{J1505$+$0326: The apparent long rotation. With the exception of the left-most point each angle measurement (solid symbols) is paired with its $180^\circ$ conjugate (empty symbols). The upper values note EVPA differences ($\Delta\chi$) and the lower ones their uncertainties ($\sigma_{\Delta\chi}$). Red marks highlight points in which the uncertainty $\sigma_{\Delta\chi}$ in $\Delta\chi$ is so large that both solutions of Eq.~\ref{eq:chi}, $\chi$ and $ \chi + \pi$, could be valid making the direction of the rotation uncertain. 
}
\label{fig:1505_rotation}
\end{figure*}

Figure~\ref{fig:1505_rotation} zooms on the largest of the potential rotations (MJD 56790 -- 56880). Its change in angle $\Delta \chi$ is $-309.5^\circ$ and lasts for approximately 83 days yielding a mean rotation rate of $-3.7$~deg~d$^{-1}$. However, the combination of sparse sampling and large uncertainties in the angle makes the estimate of the {direction of a rotation highly uncertain which makes the detection of the rotation itself uncertain}.

As we discussed in Sect.~\ref{sec:evpa_monecla}, for each pair $(q, u)$ we choose the solution of Eq.~\ref{eq:chi} for which the absolute difference, $|\Delta\chi|$, from the previous angle is less than $90^\circ$. This condition controls the direction of the EVPA evolution. However, the uncertainty associated with each angle computation must also be accounted for when this condition is checked. If it  happens that the sum of the absolute difference  $|\Delta\chi|$ between two consecutive data points and the uncertainty in that difference $\sigma_{\Delta\chi}$ 
  exceeds $90^\circ$ i.e. $90^\circ\le |\Delta\chi| + \sigma_{\Delta\chi}$, the direction of the rotation becomes uncertain as both solutions of Eq.~\ref{eq:chi}, $\chi$ and $ \chi + \pi$ could be valid\footnote{Clearly, in the absence of physical constrains any solution of the form $n\cdot\pi$ is equally valid. The choice of the smallest step is justified by the assumption of minimal variability.}.
     
With the exception of the earliest measurement (left-most point), each angle measurement (solid symbols) in Fig.~\ref{fig:1505_rotation} is paired with its $180^\circ$ conjugate (empty symbols). 
 The critical steps with $90^\circ\le |\Delta\chi| + \sigma_{\Delta\chi}$, are shown in red. Clearly, their  number prevents us from reliably telling the direction that the EVPA follows  making the detection of the rotation uncertain. 

\begin{figure}[h] 
\centering
\includegraphics[trim={0 0 0 0},clip, width=0.5\textwidth]{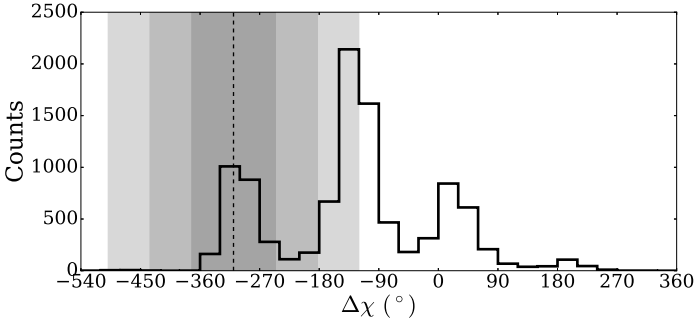} 
\caption{J1505$+$0326: The distribution of $\Delta \chi$ in simulated EVPA curves. The grey areas mark the 1, 2 and $3\sigma$ intervals while the dashed line is the observed rotation of $309.5^\circ$. The most probable value (peak of the solid distribution) is around $-140$~deg.}
\label{fig:1505_hist_Dx}
\end{figure}

The uncertainty in the detected rotation can also be shown by examining the effect of the uncertainties in $q$ and $u$ on the rotation angle. For simplicity, we  assume that the measured $q$ and $u$ are the means of the Gaussian distributed fractional Stokes parameters which is equivalent to saying that they describe the ``real" intrinsic behaviour of the source.  
We then add Gaussian noise based on their uncertainties and re-calculate the EVPA curve and compare its parameters with those of the observed one.   
In Fig.~\ref{fig:1505_hist_Dx} we show the distribution of rotation angles, $\Delta \chi$, for a total of $10^4$ simulated light curves. On the basis of our assumptions the probability of detecting a rotation with an absolute angle $|\Delta \chi|$ within $1\sigma$ of the observed value is approximately 0.22. For larger rotations ($|\Delta \chi|\ge 309.5^\circ$) the probability is around 0.081. 

From this we conclude that we cannot be confident about the intrinsic evolution of the EVPA. Even if we knew the intrinsic variability, the limited sampling and the measurement uncertainties would allow a vast range of possible EVPA curves {that would result in varying changes of the EVPA (i.e. $\Delta\chi$)}.
Subsequently, the previous test can tell us what is the most likely observation, but it cannot tell us anything about the intrinsic variability. For example, although the bin with the largest probability appears around $-140^\circ$, this does not imply that the intrinsic EVPA rotation covers, most probably, $140^\circ$. 

Finally, the data points in Fig.~\ref{fig:1505_rotation} could be aligned with roughly the same rotation rate if 180-degree shifts were chosen accordingly instead of obeying the convention of smallest change between consecutive measurements. This would result in a rotation $360^\circ$ larger than shown in Fig.~\ref{fig:1505_rotation}. 

{Clearly, the measurement of an intrinsic rotation is limited by the sparse sampling and the 180-degree ambiguity. In order to assess the reliability of the observed event we take two steps:  
\begin{enumerate}
\item We first estimate the probability that the measurement uncertainties induce a fake rotation in the absence of a real one,  
\item We estimate the likelihood of an intrinsic rotation given the observed data. 
\end{enumerate}
{The} following simulations use exactly the same time sampling as the data and thus are affected by the 180-degree ambiguity in the same way.}

\subsubsection{Is the observed rotation an artefact of noise?}
\label{subsubsect:noise}
Here we assess the probability of the observational noise inducing the apparent rotation in the absence of an intrinsic rotation; that is, assuming $d\chi_\mathrm{intr}/dt=0$~deg~d$^{-1}$. 

\begin{figure}[h] 
\centering
\includegraphics[trim={0 0 0 0},clip, width=0.5\textwidth]{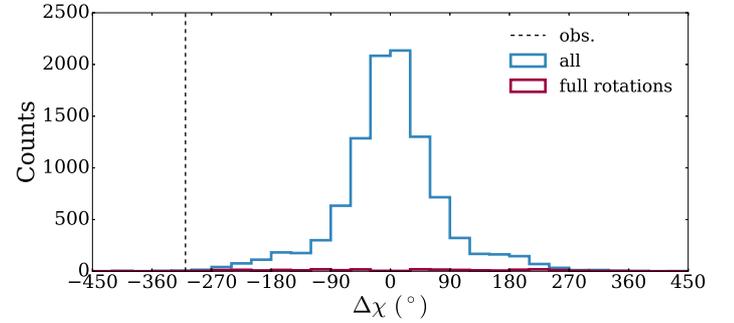} 
\caption{J1505$+$0326: The distribution of $\Delta \chi$ in the simulated EVPA curves where we assume the absence of intrinsic variability and the operation solely of noise. The dashed line is the observed rotation angle of $309.5^\circ$.}
\label{fig:1505_hist_Dx_noise}
\end{figure}

For simplicity, we set $q$ to the mean polarisation fraction in our simulations during the observed rotation $\overline{p_\mathrm{rot}}$ and $u$ to zero, which results in $\chi = 0^\circ$. Subsequently, we add Gaussian noise $\mathcal{N}$ to these values according to the estimated uncertainties:
\begin{eqnarray}
q&=&\overline{p_\mathrm{rot}} + \mathcal{N}(0, \sigma_q)\\
u&=&\mathcal{N}(0, \sigma_u).
\end{eqnarray}
$\mathcal{N}(0, \sigma)$ denotes that the noise centres at 0. We run $10^4$ simulations. For each run the same algorithm used for the observed \allbold{data was used to identify \textit{full rotations}. We define as ``full rotation'' in our simulations a rotation that consists of as many data points as the observed long rotations. 
}. The probability of finding a full rotation is:
\begin{equation}
	\label{eq:J1505_P1}
P\left(\mathrm{full~rotation}~|~d\chi_\mathrm{intr}/dt=0 \right)=2.7\times10^{-2}.
\end{equation}
We also find that 
\begin{equation}
	\label{eq:J1505_P2}
P\left(|\Delta \chi_\mathrm{intr}|\ge 309.5^\circ|~d\chi_\mathrm{intr}/dt=0 \right)=10^{-3}
\end{equation}
and
\begin{equation}
	\label{eq:J1505_P3}
P\left(\mathrm{full~rotation;}~|\Delta \chi_\mathrm{intr}|\ge 309.5^\circ|~d\chi_\mathrm{intr}/dt=0 \right)=6\times10^{-4}
\end{equation}
In Fig.~\ref{fig:1505_hist_Dx_noise} we show the results of the simulations. 

This exercise shows that although it is not impossible that the observed event is merely an artefact of noise, it is fairly improbable. Thus, there must be intrinsic variability even if we cannot be sure of its exact nature.


\subsubsection{The most probable parameters of the intrinsic event}
\label{subsubsect:1505_mostProb}
{Having shown that intrinsic variability seems much more likely to be driving the observed EVPA behaviour, we wish to estimate the most probable parameters of the potential intrinsic rotation.}

\begin{figure}[h] 
\centering
\includegraphics[trim={0 0 0 0},clip, width=0.5\textwidth]{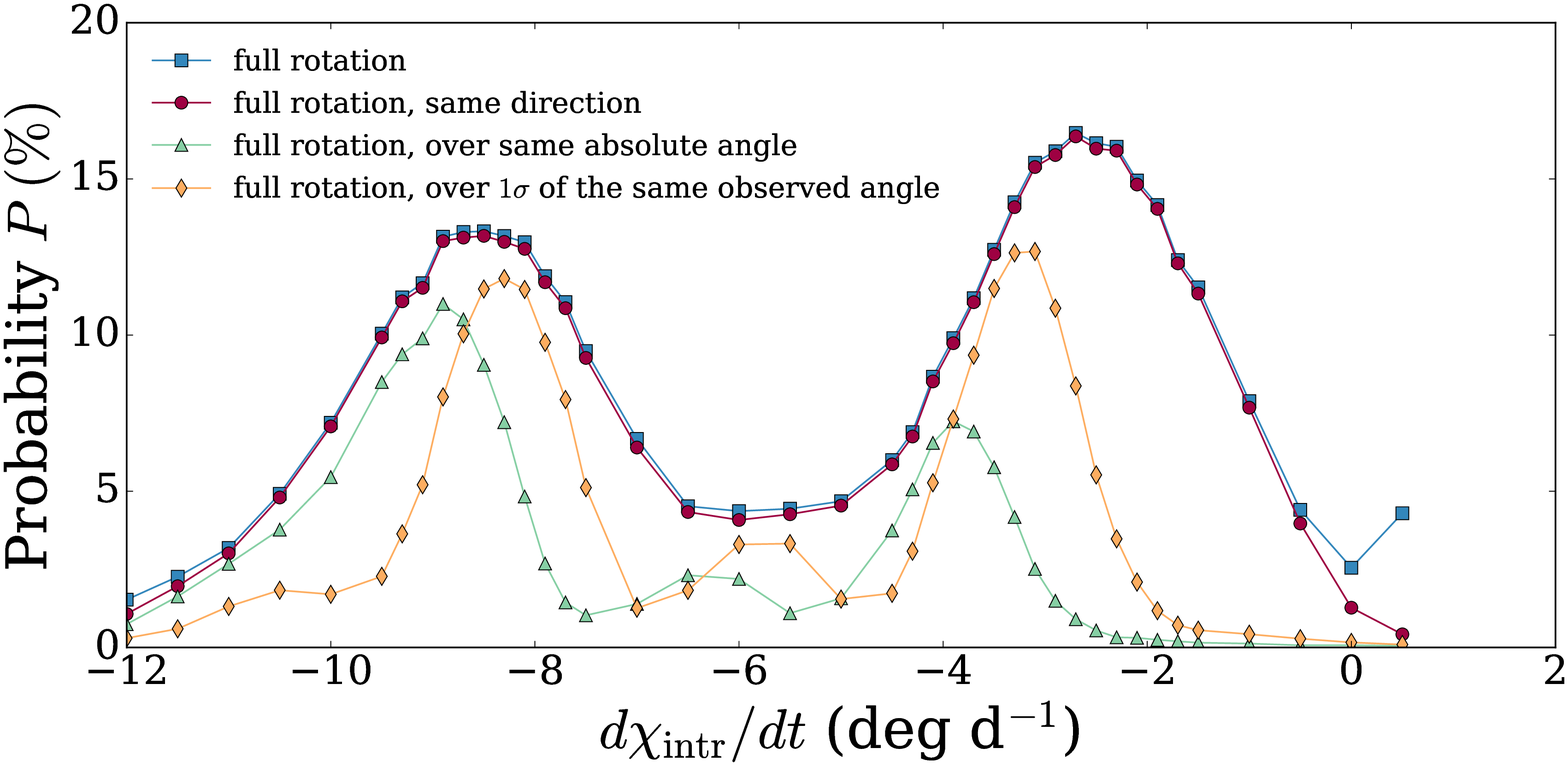} 
\caption{J1505$+$0326: The probability distributions assuming an intrinsic EVPA rotation with a constant rate.}
\label{fig:1505_probaility}
\end{figure}

Our analysis relies on the assumption of a constant intrinsic rotation rate $d\chi_\mathrm{intr}/dt$ {as well as a constant polarisation fraction (during the rotation)}. The rotation is simulated in $q$--$u$ space by adding Gaussian noise to $q$ and $u$.
 We test a range of rotation rates $d\chi_\mathrm{intr}/dt=\{-12., -11.5, ... , +0.5\}$ in units of deg~d$^{-1}$. For each rate we run $25\cdot10^3$ simulations and compute the probability of: 
\begin{enumerate}
	\item observing a full rotation that is over the entire period we simulate (i.e. including all data points), blue squares in Fig.~\ref{fig:1505_probaility};
 	\item observing a full rotation in the same direction as the one in the data (in this case negative derivative in EVPA), red circles in Fig.~\ref{fig:1505_probaility}; 
 	\item observing a full rotation over an angle at least as large (in absolute terms) as the observed one ($|\Delta\chi_\mathrm{sim}| \geq |\Delta\chi_\mathrm{obs}|$ i.e. $\Delta\chi_\mathrm{sim} \leq \Delta\chi_\mathrm{obs}$), green triangles in Fig.~\ref{fig:1505_probaility};
 	\item observing a full rotation with an angle within the $1\sigma$-range of the rotation angle observed in the data ($\Delta\chi_\mathrm{obs} - \sigma_{\Delta\chi,\mathrm{obs}} \leq \Delta\chi_\mathrm{sim} \leq \Delta\chi_\mathrm{obs} + \sigma_{\Delta\chi,\mathrm{obs}}$), orange diamonds in Fig.~\ref{fig:1505_probaility}.
\end{enumerate}

%
%

Figure~\ref{fig:1505_probaility} shows the resulting probability distributions. For a full rotation over angles at least as large as the observed one the most likely intrinsic rotation rate is $-8.9\pm0.1$~deg~d$^{-1}$ with a corresponding probability of 0.11 (green diamonds). The second most probable rate is $-3.9 \pm 0.1$~deg~d$^{-1}$ with a probability of 0.068. For a full rotation over an angle within $1\sigma$ of the observed one (orange diamonds), the most probable intrinsic rotation rate is found to be $-3.1 \pm 0.1$~deg~d$^{-1}$ with  a probability of 0.129 while the second most likely one $-8.3 \pm 0.1$~deg~d$^{-1}$ with probability 0.119. 

\begin{figure}[h] 
\centering
\includegraphics[trim={0 0 0 0},clip, width=0.5\textwidth]{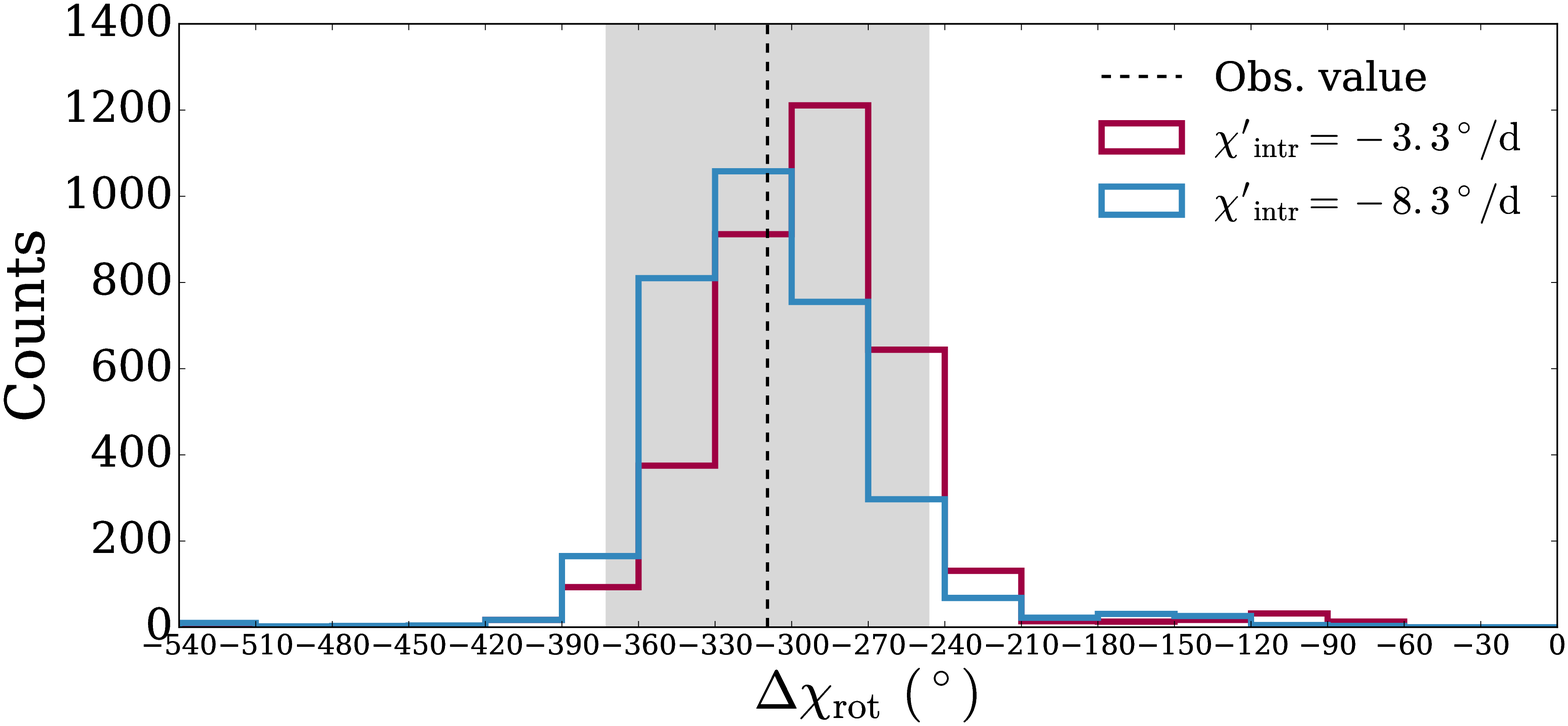} 
\includegraphics[trim={0 0 0 0},clip, width=0.5\textwidth]{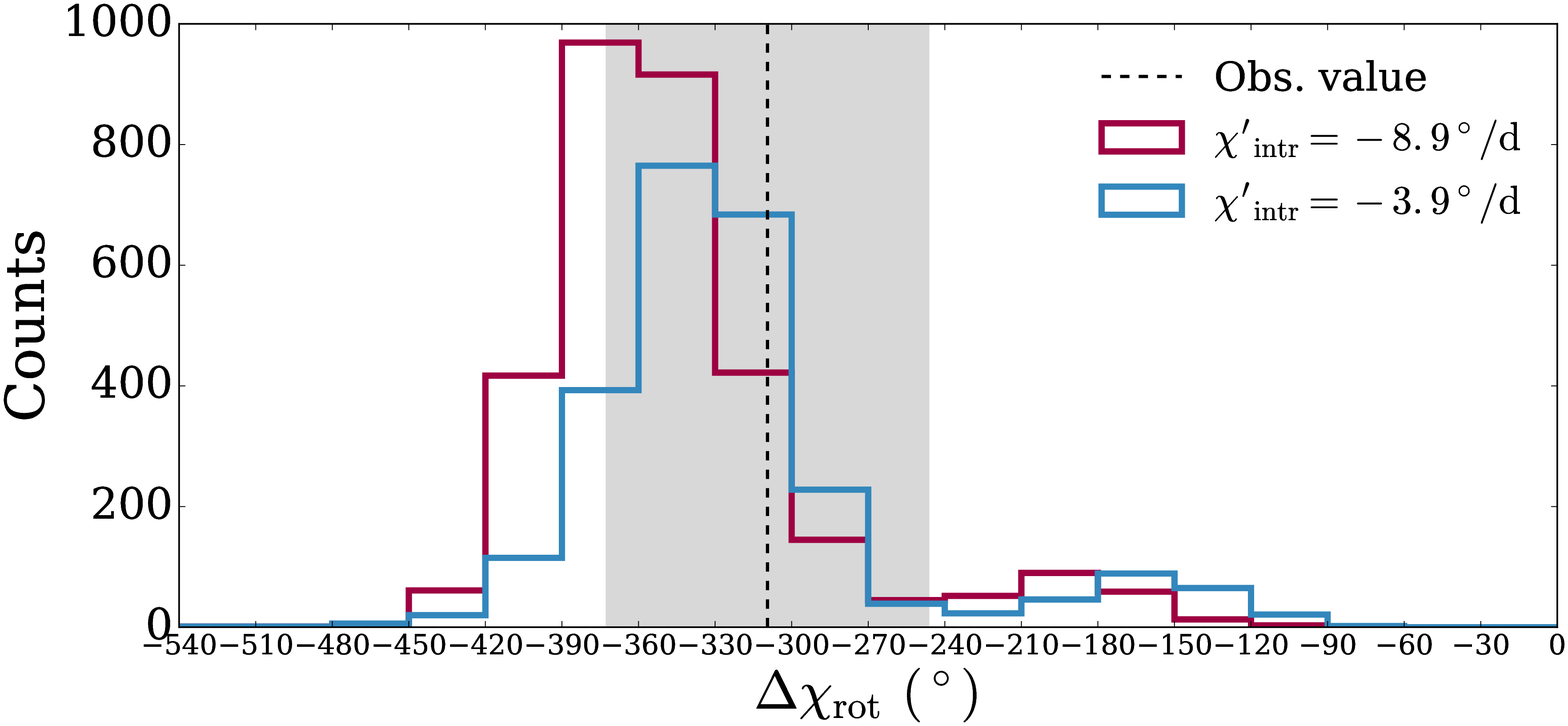} 
\caption{J1505$+$0326: The distribution of rotation angles at the most likely intrinsic rotation rates based on the $1\sigma$ (upper panel) and the extreme-span-criterion (lower panel, case 3 of Section~\ref{subsubsect:1505_mostProb}). The dashed line marks the observed value and the grey area the $1\sigma$ uncertainty.}
\label{fig:1505_DistAmplitude}
\end{figure}

In Fig.~\ref{fig:1505_DistAmplitude} we show the distribution of rotation angle at the most likely intrinsic rotation rates based on the $1\sigma$-criterion (upper panel) and on the basis of the extreme-span-criterion (lower panel, case 3 of Section~\ref{subsubsect:1505_mostProb}). The observed rotation angle (dashed vertical line) is consistent with an intrinsically constant rotation.

{Considering all this and the conclusions of Section~\ref{subsubsect:noise}, we realise that it is much more likely that an intrinsic EVPA rotation (with the addition of pseudo-variability introduced by the uncertainties) is causing the observed event. {Assuming a constant rotation rate also shows that the observed rotation is more likely the result of an intrinsic event rather than that of pure noise.} 

\subsection{J0324$+$3410}
\label{subsec:J0324}

In Fig.~\ref{fig:0324_t_p_evpa} we show the de-biased polarisation fraction $\hat{p}$ and angle $\chi$ as a function of time. The dataset includes RoboPol, KANATA, Perkins, and the Steward observatory measurements. 
\begin{figure*}[] 
\centering
\includegraphics[trim={0 0 0 0},clip, width=0.9\textwidth]{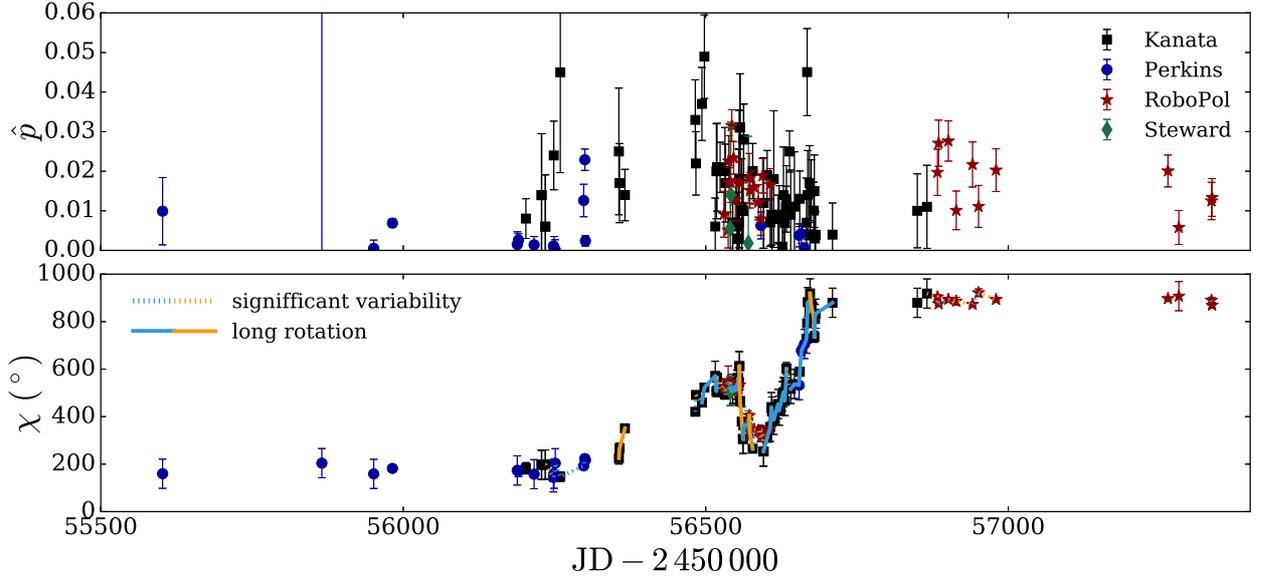} 
\caption{J0324$+$3410: The polarisation variability curve. \textit{Upper panel:} The de-biased polarisation fraction over time. \textit{Lower panel:} The EVPA over time. The coloured lines mark periods of monotonous -- within the uncertainties -- EVPA evolution.}
\label{fig:0324_t_p_evpa}
\end{figure*}
{Both $\chi$ and $p$ show significant variability.} 

{The polarisation fraction $\hat{p}$ spreads around a median of 0.012 with a standard deviation of 0.016 (see Fig.~\ref{fig:J0324_pol_hist} and Table~\ref{tbl:cumm_pol}).}
{The distribution of the EVPA confined in the $\left[-90\degr,90\degr\right]$ range is shown in Fig.~\ref{fig:J0324_EVPA_HIST}. It distributes rather narrowly around a median of $-6.7\degr$ (with standard deviation of $40.2\degr$). This preferred direction is at an angle of $49.3\degr$ with the 15-GHz radio jet axis which found to be remarkably stable at a position angle of $124\degr$  \citep{2016RAA....16..176F}.}

{In total, we detected 28 apparent rotations of the polarisation plane with rotation angle $\Delta\chi$, ranging from approximately $19\degr$ to $402\degr$ (Fig.~\ref{fig:0324_hist_dchi}). Ten of those qualify as long rotations as they include at least three measurements and exceed $90^{\circ}$. In Fig.~\ref{fig:J0324_pol_hist} we show the distribution of $\hat{p}$ during phases of rotation and of non-rotation separately. In the former case the median $\hat{p}$ is 0.014 and in the latter only 0.007. This indication that the polarisation fraction centres around different values in those two phases is not supported by a two-sample KS.}
\begin{figure}[] 
\centering
\includegraphics[trim={0 0 0 0},clip, width=0.5\textwidth]{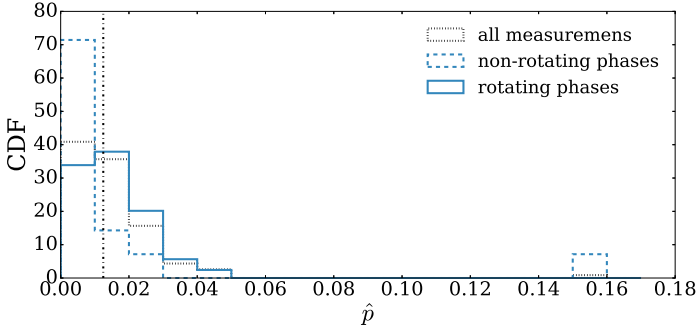} 
\caption{{J0324$+$3410: The distribution of the de-biased polarisation fraction $p$. The dot-dashed vertical line marks the median of the distribution. The blue dashed line shows the distribution of $\hat{p}$ during the non-rotating phases and the solid one that during the rotating phases. The black dotted line corresponds to all the measurements. }}
\label{fig:J0324_pol_hist}
\end{figure}
\begin{figure}[] 
\centering
\includegraphics[trim={0 0 0 0},clip, width=0.5\textwidth]{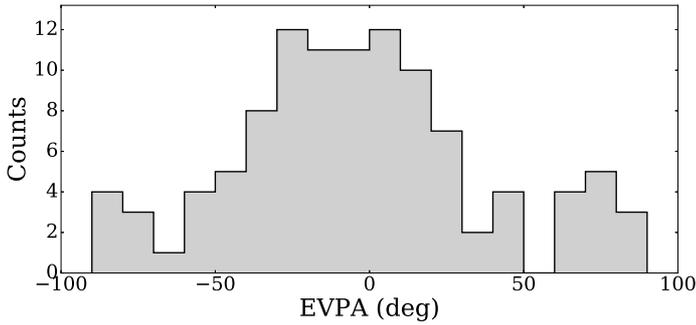} 
\caption{{J0324$+$3410: The distribution of the observed polarisation angles in the range $\left[-90\degr,90\degr\right]$. }}
\label{fig:J0324_EVPA_HIST}
\end{figure}
\begin{figure}[] 
\centering
\includegraphics[trim={0 0 0 0},clip, width=0.5\textwidth]{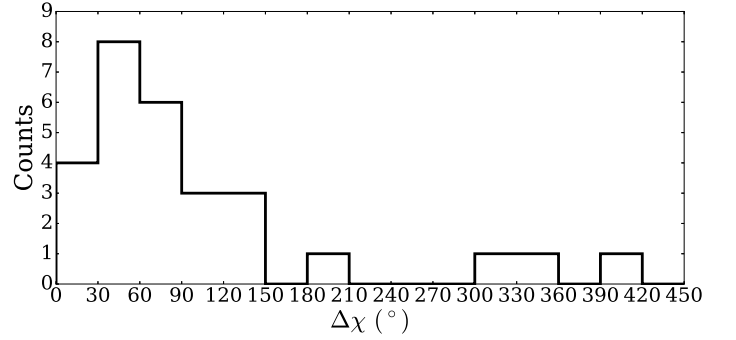} 
\caption{{J0324$+$3410: The distribution of rotation angles $\Delta\chi$ for all the apparent rotations.}}
\label{fig:0324_hist_dchi}
\end{figure}

{As in the case of 1505$+$0326, in the following we concentrate on the rotation candidates}. Figure~\ref{fig:0324_t_p_evpa_long} focuses on the area where two major potential rotations occur. The largest one happens at around MJD 56640.5 -- 56672.4 over $402\pm87^\circ$ corresponding to a mean rate of approximately $13$~deg~d$^{-1}$. The second largest event happens around MJD 56595.6 -- 56633.5 with an angle of $349\pm66^\circ$ and a mean rate of $9$~deg~d$^{-1}$.

\subsubsection{The largest potential rotation}

Figure~\ref{fig:0324_rotation} (upper panel) demonstrates the uncertainty associated with the evolution of the measured EVPA. All steps are critical and one cannot be certain of the direction the EVPA intrinsically takes at any point in its evolution. {Thus, making the very detection of the rotation uncertain.} 

Following the approach presented in Section~\ref{subsubsect:noise}, we examine whether the uncertainties in $q$ and $u$ alone can cause the  observed rotation in the absence of an intrinsic rotation. We assume again that the measured  $q$ and $u$ are correct estimates of the means of the Gaussian distributed Stokes parameters. After running $10^4$ simulations we find that the probability of finding one full rotation (passing over all points) is $\sim2\times10^{-2}$ (Eq.~\ref{eq:J1505_P1}) while that of finding a full rotation with absolute angle larger than observed, only $8\times10^{-4}$ (Eq.~\ref{eq:J1505_P3}). {Hence, although it is not impossible that the observed event is an artefact of noise while the EVPA remains intrinsically unchanged, it is rather unlikely. The associated probability is only $\sim10^{-3}$. It is then possible that the EVPA indeed undergoes an intrinsic variability event.} 
\begin{figure*}[] 
\centering
\includegraphics[trim={0 0 0 0},clip, width=0.9\textwidth]{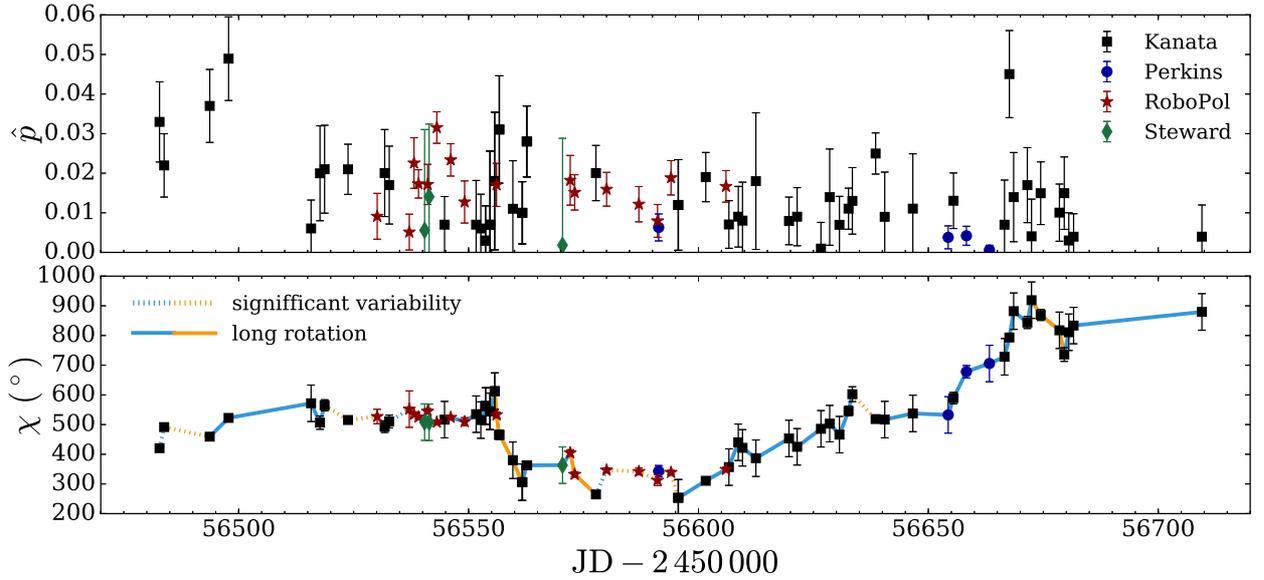} \\
\caption{J0324$+$3410: A zoom on the region of the largest potential rotations.}
\label{fig:0324_t_p_evpa_long}
\end{figure*}
\begin{figure*}[h] 
\centering
\includegraphics[trim={0 0 0 0},clip, width=0.9\textwidth]{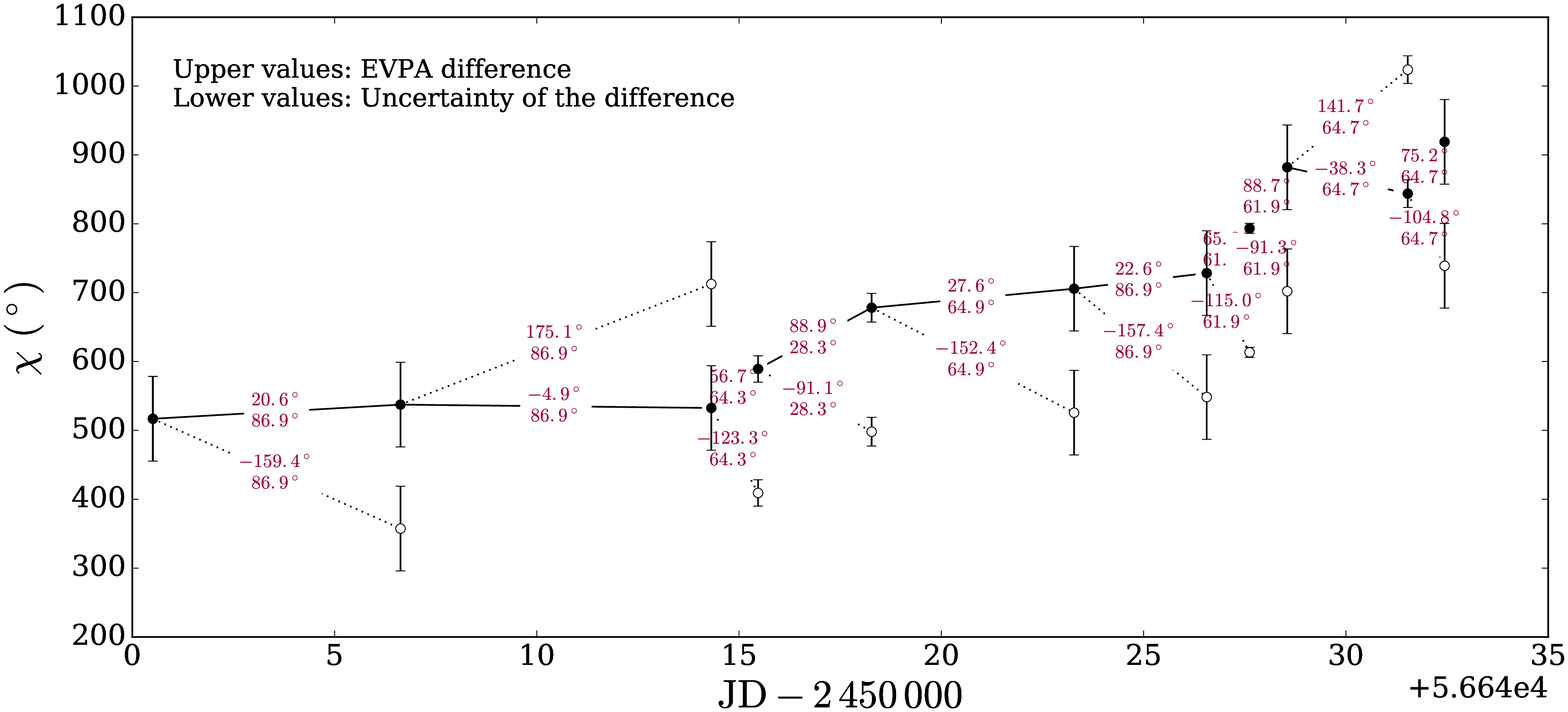} 
\includegraphics[trim={0 0 0 0},clip, width=0.9\textwidth]{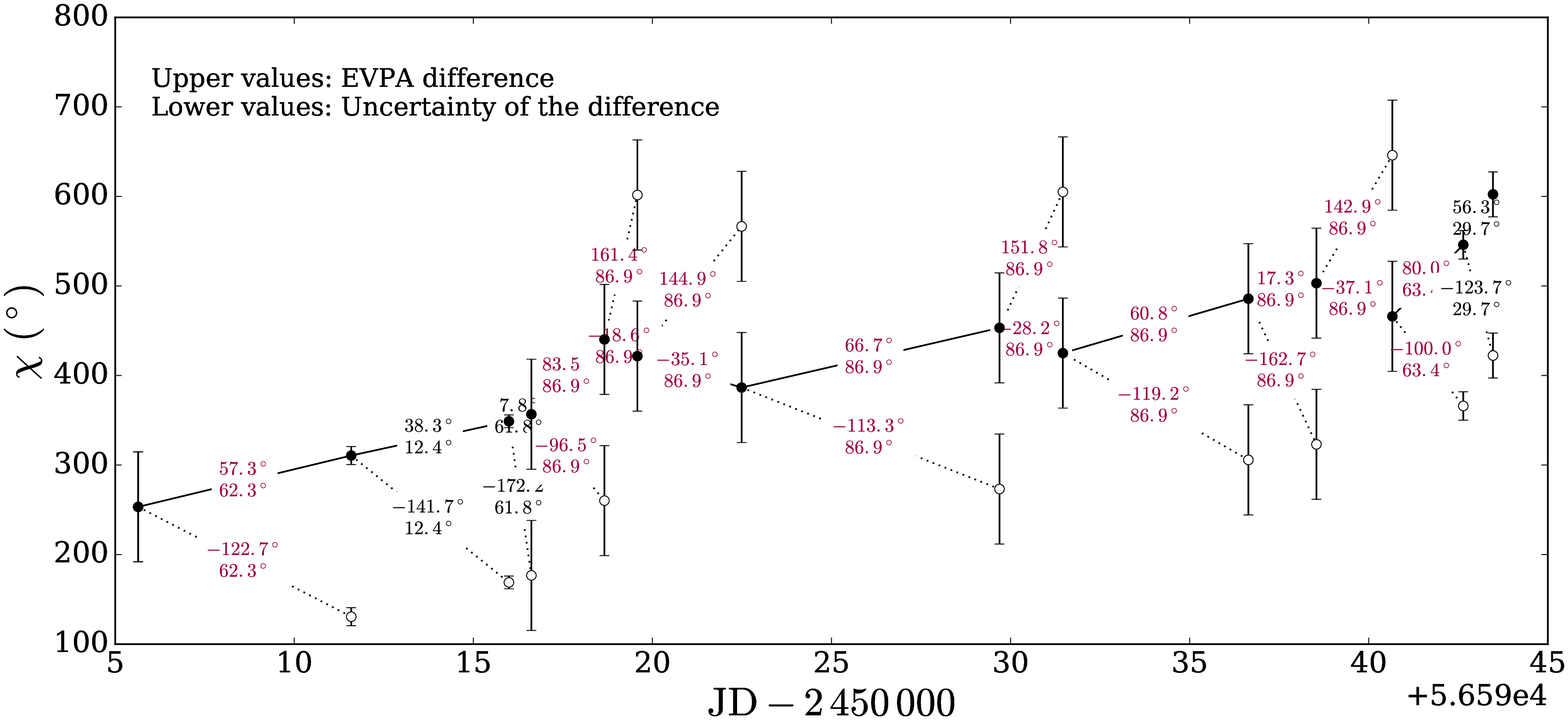} 
\caption{J0324$+$3410: 	The two major potential EVPA rotations: the largest potential (\textit{upper} panel) and the second largest (\textit{lower} panel). The uncertainty in the observed events caused by sparse sampling and large measurement uncertainties becomes apparent.}
\label{fig:0324_rotation}
\end{figure*}

To estimate the most probable parameters of the intrinsic EVPA variability, we make the assumptions {(as in Sect.~\ref{subsubsect:1505_mostProb}) of constant intrinsic EVPA rotation rate, and the constancy of the polarisation fraction during the intrinsic rotation}.
After $2.5\times10^4$ iterations we find that the most likely intrinsic rotation rate for a full rotation with an angle at least as large as the observed one, is $19\pm0.5$~deg~d$^{-1}$ {(probability is about $10^{-2}$) }. The most probable rotation rate for a full rotation with an angle within $1\sigma$ of the observed one, is $10\pm0.25$~deg~d$^{-1}$ (probability is about 0.033). 

{These probabilities are indeed {low, yet, they} are higher than those for the pure noise scenario, indicating that intrinsic variability is more likely. Nevertheless, the low probability indicates that the simple assumption of a constant rotation rate is anyhow not likely.}

\subsubsection{The second largest potential rotation}
In the lower panel of Fig.~\ref{fig:0324_rotation} we show the second largest potential rotation. The probability that this event is the mere result of noise is as low as $\sim10^{-3}$. 
We repeat the analysis already presented earlier. After $2.5\cdot10^4$ simulated EVPA curves we find that for a full rotation with an angle at least as large as the observed one, the most probable intrinsic rate is $10\pm0.25$~deg~d$^{-1}$ ($P=0.01$). For a full rotation over an angle within $1\sigma$ of the observed one, the most probable rate is $9.5\pm0.25$~deg~d$^{-1}$ ($P=0.008$). 
 
{These probabilities are low and comparable to those of the noise artefact hypothesis. An intrinsic rotation of constant rate is only marginally more likely than noise. A realistic scenario would be that the observed behaviour results from the combination of intrinsic variability and observational noise. The noise, however,	 makes the observed rotation angle an inadequate indicator of the intrinsic behaviour. Hence, although there may be intrinsic variability, we cannot recover it due to the noisy data.}

\subsection{{J0849$+$5108 and J0948$+$0022: Polarisation Variability}}
In Fig.~\ref{fig:0849_0948_t_p_evpa} we show the $\chi$ and $\hat{p}$ datasets for J0849$+$5108 (\textit{upper} panel) and J0948$+$0022 (\textit{lower} panel) as a function of time. 
{As shown in Fig.~\ref{fig:0849_0948_t_p_evpa} and summarised in Table~\ref{tbl:cumm_pol}, the limited dataset does not allow sound quantification of the variability characteristics for either of the polarisation parameters. Intense variability is, nevertheless, clearly visible for both $\hat{p}$ and $\chi$.} 

{For J0849$+$5108 the median $\left<\hat{p}\right>$ is around 0.1 with a standard deviation 0.078. Concerning the angle, the available dataset revealed a total of nine rotations none of which exceeded $90^\circ$.} Clearly, despite the clear signs of variability, the data sparseness prevents any understanding of the intrinsic nature of the variability.  

In the case of J0948$+$0022 the slightly richer dataset (Kanata, Perkins, RoboPol and Steward) reveals the occurrence of 11 rotations two of which over angles beyond $90^\circ$. For the largest rotation the EVPA changed by $268^\circ$ (Fig.~\ref{fig:0948_rotation}). Although this EVPA curve is fairly reliable the steps $\chi_\mathrm{i}-\chi_\mathrm{i+1}$ between adjacent data points are very large and close to 90 degrees making this curve unsuitable for further analysis. 
Subsequently, nothing can be said as to whether the source underwent intrinsic EVPA rotations or not. For both sources better sampled datasets are necessary. {As we did in previous sections, in Fig.~\ref{fig:J0948_pol_hist} we show the cumulative distribution function of $\hat{p}$ separately for all measurements and for phases with rotation candidates. The median $\left<\hat{p}\right>$ is around 0.024 with a spread of 0.028. During phases of rotation the median is 0.028 and over non-rotating phases it drops to 0.016. As in previous cases, however, a two sample KS test did not support the hypothesis of different behaviour over the two states of activity.}
\begin{figure*}[h] 
\centering
\includegraphics[trim={0 0 0 0},clip, width=0.9\textwidth]{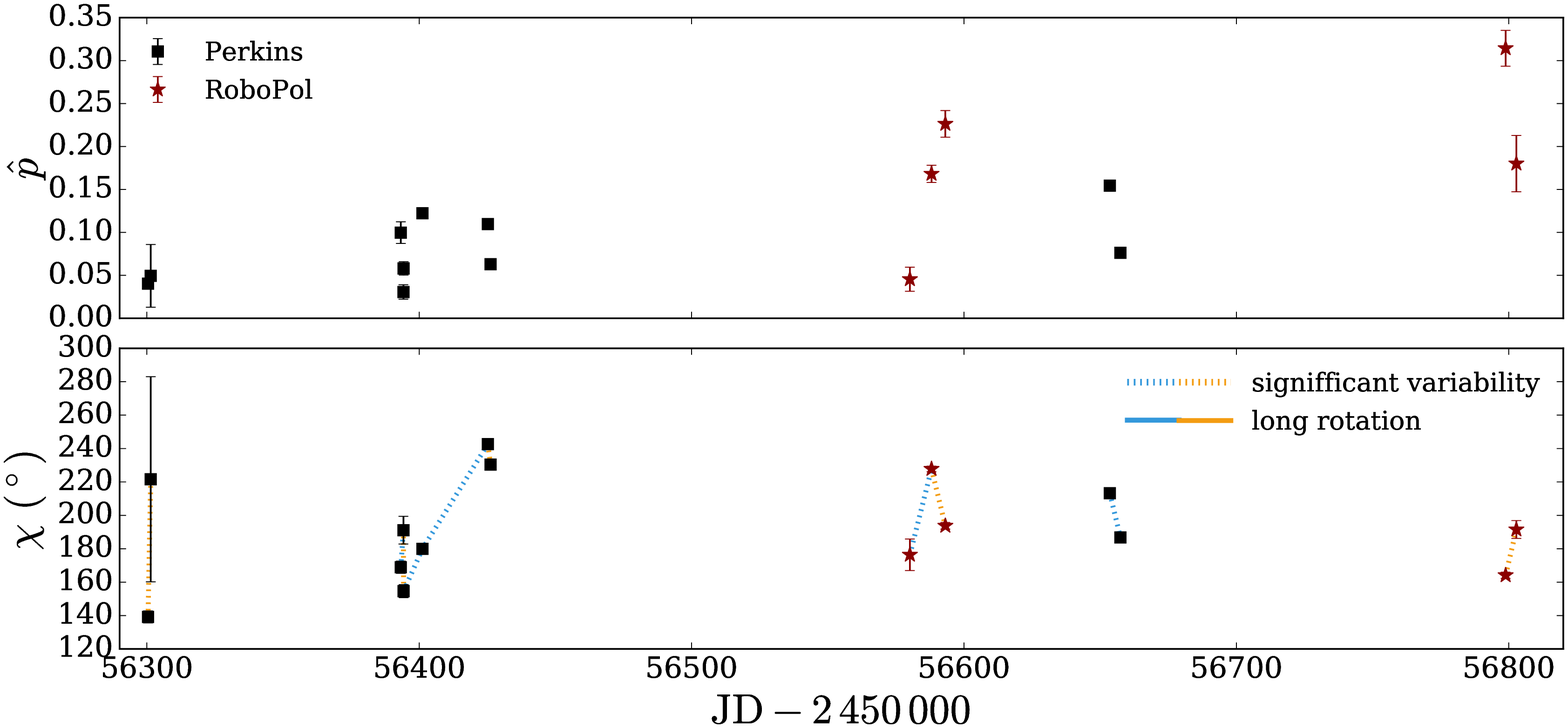} 
\includegraphics[trim={0 0 0 0},clip, width=0.9\textwidth]{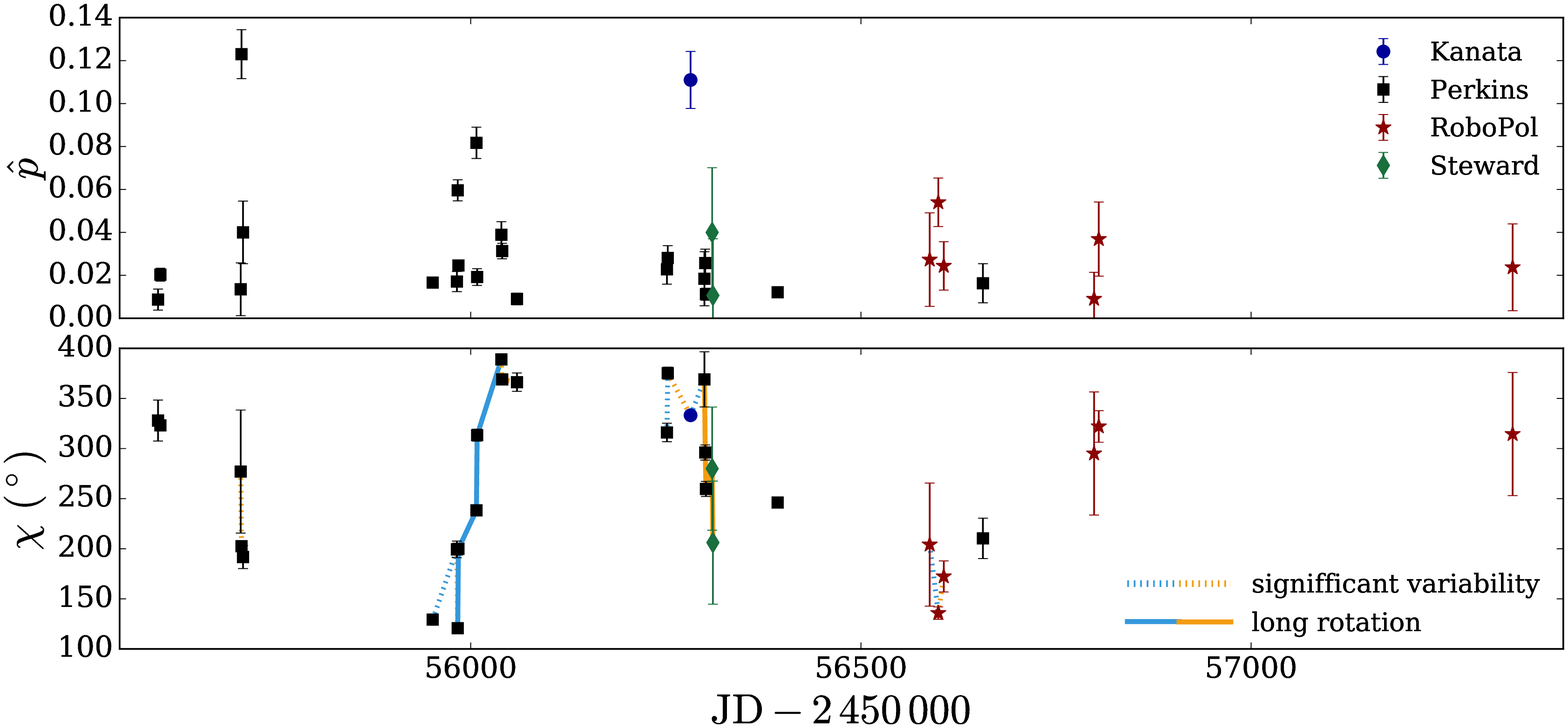} 
\caption{Polarisation fraction and EVPA as a function of time for J0849$+$5108 (\textit{upper} panel) and J0948$+$0022 (\textit{lower} panel). The coloured lines mark periods of significant, monotonous -- within the uncertainties -- EVPA evolution. 
Dotted lines mark periods of significant evolution while solid ones periods of long rotations.}
\label{fig:0849_0948_t_p_evpa}
\end{figure*}
\begin{figure*}[h] 
\centering
\includegraphics[trim={0 0 0 0},clip, width=0.9\textwidth]{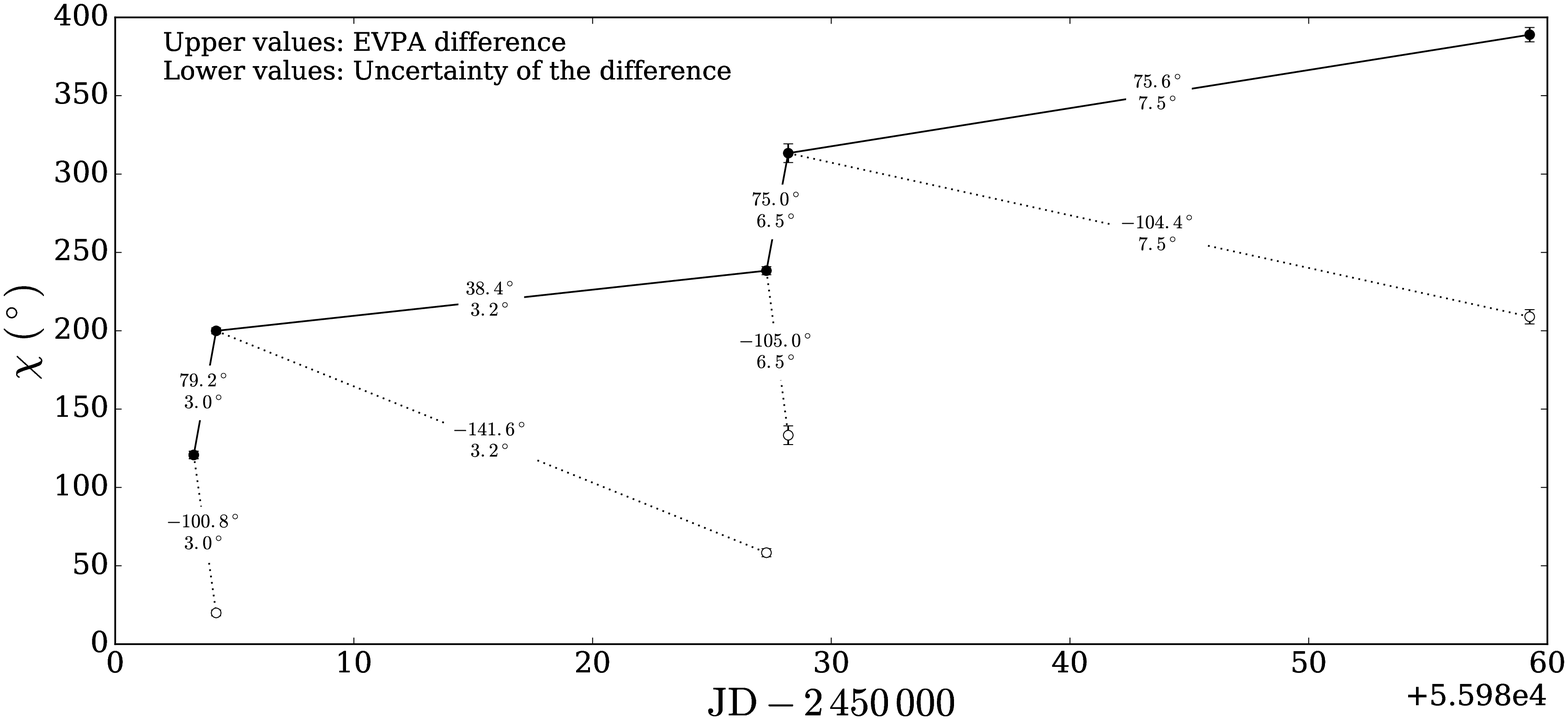} 
\caption{J0948$+$0022: the major potential EVPA rotation. The uncertainty in the observed events caused by sparse sampling and large uncertainties is obvious.}
\label{fig:0948_rotation}
\end{figure*}
\begin{figure}[h] 
\centering
\includegraphics[trim={0 0 0 0},clip, width=0.5\textwidth]{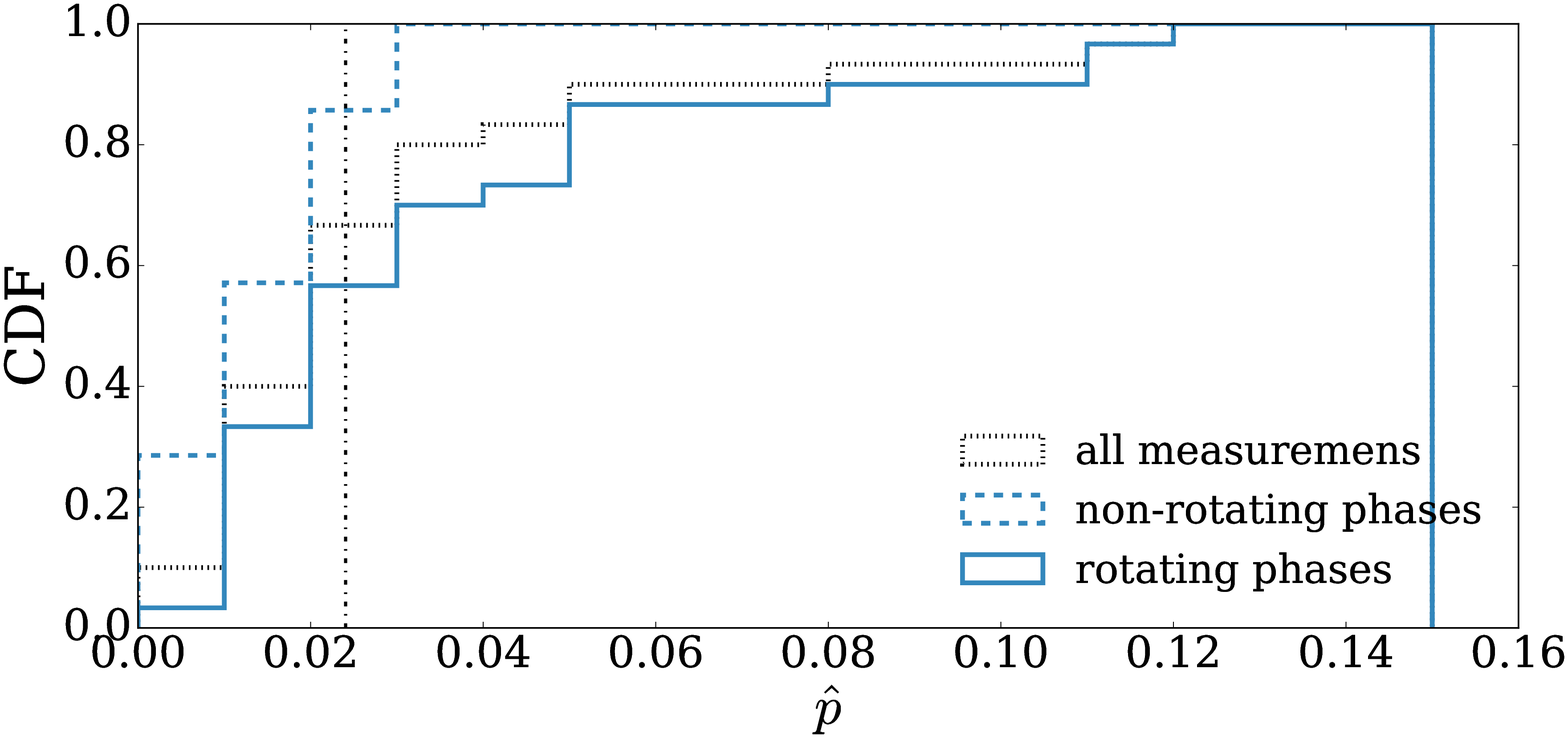} 
\caption{{J0948$+$0022: The distribution of the de-biased polarisation fraction $\hat{p}$. The dot-dashed vertical line marks the median of the distribution. The blue dashed line shows the distribution of $\hat{p}$ during the non-rotating phases and the solid one that during the rotating phases. The black dotted line corresponds to all the measurements. } }
\label{fig:J0948_pol_hist}
\end{figure}
\section{Discussion}

\allbold{As a result of the small availability of high cadence datasets, the current understanding of the optical polarisation variability of RL~NLSy1s is only sparse and has been gained by selected case studies rather than a systematic population study. PMN\,J0948$+$0022, for example, was found to have variable optical polarisation  \citep[degree and angle, ][]{2012AAS...21924329E,2013ApJ...773...85E} even on minute time scales \citep{2013ApJ...775L..26I}. The latter reported that the polarisation briefly exceeded  30~\% while the EVPA remained unchanged. \cite{2014ApJ...794...93M} reported that J0849$+$5108 also showed rapid, intra-night variability both in degree of polarisation and angle. 1H\,0323$+$342 was found by \cite{2014PASJ...66..108I} to possess an EVPA that remained constantly parallel to the jet orientation. All these indicate how unknown the optical polarisation from such systems and its temporal behaviour remains. Our study attempts to overcome this barrier by examining a larger sample that would potentially allow us to extract general conclusions and further compare them with blazars. Larger systematic studies are, however, needed to further clarify: (a) the confirmation of EVPA rotation events in RL~NLSy1s and later a comparison of their parameters to those in blazars, (b) the behaviour of the polarisation degree during rotations of the EVPA  \citep{2016MNRAS.457.2252B}, and (c) the association of rotation events to the GeV energy band activity \citep{2018MNRAS.474.1296B}. Below we discuss some of the understanding we have gained with this work which seem to indicate similarities to the behaviour seen in blazars.}  
\\\\
\textit{The polarisation variability.} As we have shown, the optical polarisation parameters of RL~NLSy1 galaxies show clear signs of variability. {Both the polarisation fraction $\hat{p}$ and angle $\chi$ show phenomenologies similar to those seen} in blazars, indicating similar processes in the two classes. 
{It would be premature to conclude that the observed variability of RL NLSY1s implies turbulent processes until much higher cadence data sets of a much larger sample of sources has been considered. }
The ``blazars-like'' character of these sources however is established anew and it is important for future studies to focus on examining the fundamental differences and similarities between the two classes.  
\\\\
{\textit{The polarisation angle variability.} The very detection of significant variability of the EVPA, represents a particularly important finding. Earlier studies have indicated that -- in  selected cases -- the polarisation angle remained stable even during phases of intense polarisation fraction variability. This stability was interpreted as an indication for a high degree of magnetic field organisation in the regions where the radiation is produced. Our analysis establishes that the variability of the polarisation angle is common in all our sources with at least moderate sampling.} 
\\\\
\textit{Preferred orientation of the polarisation plane.} In the one case with a sufficiently large dataset (namely J0324$+$3410) we examined the distribution of the polarisation plane orientations. This was done by confining the polarisation orientations within the range $\left[-90\degr,90\degr\right]$. The EVPA of this source is not oriented randomly. It instead, shows a concentration (with some breadth) around a preferred direction of $-6.7\degr$. This orientation is at $49.3\degr$ to the position angle of the 15-GHz radio jet. The behaviour of preferential orientation of the EVPA resembles the high synchrotron peak frequency sources discussed by \cite{2016MNRAS.463.3365A,2017arXiv171104824A}.

Under the assumption that the optical emission is thin, this configuration implies a projected net magnetic field that is oriented at $40.7\degr$ to the radio jet\footnote{In the optically thin regime of synchrotron emission the projected magnetic field is perpendicular to the observed EVPA.}. 
This misalignment could be an indication of a combination of poloidal and toroidal field components. A more systematic approach, of course, would require the careful consideration of projection effects, the location and size of the emission region (potentially very different for the optical and radio emission), and the effects of relativistic aberration \citep{2005MNRAS.360..869L}. The variability accompanying the preferential orientation, on the other hand, could be understood in terms of an additional turbulent magnetic field component \citep[e.g.][]{2014ApJ...780...87M}, a regular modulation of the emission region location \citep[e.g.][]{2008Natur.452..966M}, and relativistic delays and projection effects \citep{2014ApJ...789...66Z}. Such configurations are not un-realistic. It should be noted, however, that the optical emission is likely originating from a more compact region much closer to the jet base when interpreting these observations.  
\\\\
{\textit{The detection of long polarisation plane rotations.} In the general framework of polarisation variability we have searched for long rotations of the polarisation plane and assessed the probability of such events being driven by intrinsic rotations of the polarisation plane. As we have shown, the events detected (even for the best sampled EVPA curves) have to be viewed with caution. The probability that they are the mere result of noise is not null. Our analysis, however, shows clearly that the probability of an intrinsic event driving the apparent behaviour is significantly larger; especially in the case of J1505$+$0326. We emphasise the importance of this analysis for the assessment of these probabilities which we consider an equally important part of our work as the detection of the rotation events themselves. Denser datasets of larger samples are necessary for proving that the occurrence of such events is a general characteristic of RL NLSy1s.  } 
\\\\
\textit{The fractional polarisation during rotations.} As it was suggested by \cite{2016MNRAS.457.2252B}, EVPA rotations in blazars seem to be associated with lower fractional polarisation in a statistical sense. This is a rather mild effect but it may provide a diagnostic for the rotation mechanisms \citep{2017MNRAS.472.3589K}. {Although our dataset is insufficient to test this, 
for the two largest datasets (J0324$+$3410 and J1505$+$0326) the polarisation tends to be marginally higher during the rotation phases. 
A two-sample KS tests, however, did not provide any evidence that the distributions of $\hat{p}$ in the two activity phases are really different. This ambiguity will be studied in a future publication. }
\\\\
\textit{Physical interpretation.} {Clearly, the current dataset cannot shed light on the physical interpretation of the observed variability.  {We cannot tell} whether it is the physical rotation of the emission element on a helical trajectory, the macroscopic properties of the jet, turbulent processes resulting in random walks, light travel time effects or any other process that causes EVPA rotations. Much longer and better sampled light curves of larger samples are necessary for proving that apparent EVPA rotation is caused by intrinsic variability. Further studying whether EVPA variability is correlated with gamma-ray flaring might also provide insights into the physical processes driving these rotations. \cite{2018MNRAS.474.1296B} reported that in blazars there has been no EVPA rotations detected that are not associated with some activity in the \textit{Fermi} energy bands. This is of fundamental importance {to} understanding {of} the mechanism behind the long rotations of the polarisation plane. Investigating whether this is true for RL NLSy1s would be a natural next step. }
\\\\
\section{Conclusions}
\label{sec:conclusions}

We have conducted  optical polarisation monitoring of a sample of {10 RL~NLSy1} galaxies five of which have been found to radiate significant MeV -- GeV emission. {Our main goal was to quantify the variability of the two polarisation parameters, de-biased fraction $\hat{p}$ and angle $\chi$. We further examined whether long rotations of the EVPA {(similar to those found in blazars)}, are present}. Our main conclusions are as follows:  
   \begin{enumerate}
	   \item {All cases with adequately large datasets both $\hat{p}$ and $\chi$ show significant variability.}
      \item For the four {GeV emitting sources in our sample} and for which dense and long enough datasets were available, we find significant variability in the EVPA. For the remaining either the sparseness of the datasets or the noise do not allow such studies.
	  \item {In the case J0324$+$3410 we find that the EVPA spreads around a preferred orientation that is at an angle of $49.3\degr$ to the 15 GHz radio jet. Hence the projected magnetic field is at  an angle of $40.7$ to the jet axis.} 
      \item In two of those cases namely J1505$+$0326 and J0324$+$3410 we have found evidence for the presence of intrinsic EVPA rotations. {Careful numerical simulations have been conducted to assess the probability that these events are driven by intrinsic variability. }
	  \item {For the three largest apparent rotations we have assessed the likeliness of the observed rotations being the result of pure observational noise in the absence of an intrinsic event. We show that although measurement uncertainties may indeed induce such behaviours, it is rather unlikely. It appears much more likely that the observed variability is indeed driven by intrinsic rotation at a constant rate.} 
	  \item For the two major candidate events we estimate the most probable parameters for the intrinsic rotation on the basis of the constant rate assumption. We conclude that a linear trend of the intrinsic rotation is more likely than a non-varying EVPA. Most likely however a more complex situation appears more realistic. Relaxing this condition would make the probability of an intrinsic event causing the observed rotation even more probable.
	   \item {For the best sampled cases we examined the behaviour of the polarisation fraction during rotation and non-rotation periods. A two-sample KS tests indicates no significant difference between them.} 
	   \item {Our analyses shows that more observations are clearly needed for further concluding an all the topics discussed here. Although, there is evidence for long rotations of the optical polarisation plane higher cadence data of larger samples are needed.} 
   \end{enumerate}

\begin{acknowledgements}
\allbold{The authors wish to thank: the anonymous referee for the very careful examination of the manuscript and the constructive comments; Dr S. Komossa for selecting the non-\textit{Fermi}-detected RL~NLSy1s of our sample, based on the criteria given in table~\ref{tab:sample}.; the internal MPIfR referee Dr N. MacDonald for the careful multiple reading of the manuscript and the detailed comments;  Dr V. Karamanavis for the project planing, discussions and comments on the manuscript; Professor V. Pavlidou for the discussions and comments as well as the support to the Skinakas observing proposal.} The RoboPol project is a collaboration between Caltech in the USA, MPIfR in Germany, Toru\'{n} Centre for Astronomy in Poland, the University of Crete/FORTH in Greece, and IUCAA in India. Data from the Steward Observatory spectropolarimetric monitoring project were used. This program is supported by Fermi Guest Investigator grants NNX08AW56G, NNX09AU10G, NNX12AO93G, and NNX15AU81G. 
\end{acknowledgements}

\bibliographystyle{aa} 
\bibliography{/Users/mangel/work/Literature/MyBIB/References.bib} 

%
%
%
%
%
%
%
%
%

\Online

\end{document}